\documentclass[12pt]{article}
\usepackage{jheppub}
\usepackage{graphicx}
\usepackage{array}
\usepackage{amsfonts}
\usepackage{amssymb}
\usepackage[inline]{enumitem}
\usepackage{amsthm}
\usepackage{subfig}
\usepackage{multirow}
\usepackage{hyperref}
\usepackage{mathrsfs}
\usepackage{accents}
\usepackage{relsize}
\usepackage{slashed}
\usepackage{shuffle}
\usepackage{tikz}
\usepackage[T1]{fontenc} 

\usetikzlibrary{patterns,decorations.markings,decorations.pathmorphing}
\tikzset{box/.pic={\filldraw[fill=black]  (0,0) circle (2.5pt);
				   \filldraw [fill=black] (0.5,0) circle (2.5pt);
			       \draw [line width=5pt] (0,0) -- (0.5,0);}}
    \definecolor{BrickRed}{RGB}{203, 65, 84}
    \definecolor{Violet}{HTML}{EE82EE}
    \definecolor{OliveGreen}{HTML}{556B2F}
    \definecolor{RoyalBlue}{HTML}{214ED3}
    
    \hypersetup{colorlinks=true, linkcolor=RoyalBlue, citecolor=OliveGreen, filecolor=OliveGreen, urlcolor=RoyalBlue, filebordercolor={.8 .8 1}, urlbordercolor={.8 .8 0}}

 \title{Scrambling and Entangling Spinning Particles}    
\abstract{In this paper we revisit the gravitational eikonal amplitudes of two scattering spinning particles and inspect their scrambling power in the spin spaces that is quantified through the tripartite information. We found that in the non-relativistic limit and a special high-energy limit the leading contribution is a quantity that is universal and theory independent. The minimal coupling is singled out with minimal scrambling in a different high momenta limit. 
We also inspected the initial state dependence of entanglement generation and found that the spin coherent state with vanishing spin may not necessarily be the hardest to entangle. Interestingly, among a family of mixed states,  the only P-rep state there known to be the best approximation of classical mixed states was singled out as one with minimal entanglement generated.}

\author[a,b,c]{Ling-Yan Hung}
\author[b]{Kaixin Ji}
\author[d,e]{Tianheng Wang}

\affiliation[a]{State Key Laboratory of Surface Physics, Fudan University, 200433 Shanghai, China}
\affiliation[b]{Department of Physics and Center for Field Theory and Particle Physics, Fudan University, Shanghai 200433, China}
\affiliation[c]{Yau Mathematical Sciences Center (YMSC), Tsinghua University, Beijing, 100084, China}
\affiliation[d]{Institute of Theoretical Physics, Chinese Academy of Science,
55 Zhongguancun Road East, 100190 Haidian District, Beijing, China}
\affiliation[e]{Institut f\"ur Physik und IRIS Adlershof, Humboldt-Universi\"at zu Berlin,
Zum Gro{\ss}en Windkanal 6, 12489 Berlin, Germany}

\begin{document}
    
\date{\today}
\maketitle

\section{Introduction}
    
The study of gravitational-wave physics~\cite{LIGOScientific:2016aoc,LIGOScientific:2017vwq,LIGOScientific:2018mvr,LIGOScientific:2020ibl,LIGOScientific:2021usb}, in particular the dynamics of binary black holes or neutron stars interacting via classical gravity, has seen a flurry of progress both in analytical~\cite{Buonanno:1998gg,Goldberger:2004jt,Kol:2007rx,Gilmore:2008gq,Foffa:2011ub,Foffa:2016rgu,Porto:2017dgs,Blumlein:2019zku,Foffa:2019rdf,Foffa:2019yfl,Blumlein:2020pog} and numerical computations~\cite{Pretorius:2005gq,Campanelli:2005dd,Baker:2005vv}. One fruitful direction is to extract classical dynamics perturbatively from scattering amplitudes in quantum field theory~\cite{Bjerrum-Bohr:2018xdl,Cheung:2018wkq,Bern:2019nnu,Bern:2019crd,Neill:2013wsa,Cristofoli:2021vyo,Bern:2020buy,Bern:2021dqo}. Modern unitarity-based techniques allow computations of scattering amplitudes at high loop orders~\cite{Bern:1994zx,Bern:1994cg,Britto:2004nc,Bjerrum-Bohr:2013bxa,Luna:2017dtq,Kawai:1985xq,Bern:2008qj,Bern:2010ue,Bern:2012uf,Bern:2019prr} and designated effective field theories~\cite{Cheung:2018wkq,Bjerrum-Bohr:2002gqz} serve to connect quantum amplitudes and classical observables. Such classical observables~~\cite{Kosower:2018adc,Maybee:2019jus,Mogull:2020sak,Jakobsen:2021smu,Jakobsen:2021lvp,Jakobsen:2021zvh,Jakobsen:2022fcj} can also be computed from generating functions. In particular, one such generating function, the classical eikonal phase, is closely related to scattering amplitudes Fourier transformed to impact parameter space~\cite{Parra-Martinez:2020dzs,Bern:2021dqo,DiVecchia:2021bdo,Heissenberg:2021tzo,DiVecchia:2022nna}.

In addition to being the generating function for physical observables, the eikonal phase can also be viewed as the resummation of ladder diagrams in the classical approximation~\cite{Levy:1969cr,Amati:1987uf,DiVecchia:2019myk,DiVecchia:2020ymx}. 
Currently the eikonal phase has been calculated up to the third Post-Minkowskian(PM) order for non-spinning binary systems~\cite{Parra-Martinez:2020dzs,DiVecchia:2021bdo,Heissenberg:2021tzo}, and the quadratic order in spin vectors at the second PM order for spinning black holes~\cite{Jakobsen:2021zvh,Jakobsen:2022fcj}. The radiation reaction contribution to the eikonal is also studied up to the third PM order for the spinning case~\cite{Alessio:2022kwv} using the leading soft theorem~\cite{DiVecchia:2021ndb}. Moreover, the amplitude has been calculated up to the fifth order in spin at the second PM order in for spinning binary systems~\cite{Bern:2020buy,Bern:2022kto,Chen:2021qkk}, from which the 2PM eikonal phase can be extracted.\footnote{For the non-spinning case, the amplitude is available at 4PM~\cite{Bern:2021yeh}. Recently there have also been related works on the Compton scattering involving Kerr black holes at the second PM order~\cite{Aoude:2022trd,Aoude:2022thd}.}

It is curious in this semi-classical limit whether one can dig out remnants of quantum effects. One potentially fruitful line of considerations makes use of tools in quantum information theory. For example it has been suggested that should gravitational interaction increase the entanglement between interacting particles, that would certainly suggest that gravity must  not be a {\it local operation and classical communication } (LOCC)\footnote{There are controversies questioning if generation of entanglement by gravity is a sufficient smoking gun for the quantum nature of gravity. Currently it still appears possible that if for sufficiently short wavelength of the mediating gravitational wave the quantum nature of gravity might still be tested. A list of papers could be found in \cite{vedral_ans}. } \cite{Bose:2017nin, Marletto:2017kzi}.   This line of thought induced many studies of entanglement property of the S-matrix, which focused particularly on the generation of entanglement in a scattering process, for example in \cite{Cervera-Lierta:2017tdt, Beane:2018oxh, Afik:2020onf}. (For a cute use of the relative entropy see also \cite{Bose:2020shm}).  Pertaining to gravitational interaction, more recently, it was found numerically that the generation of entanglement between two scattered spinning particles is minimal at minimal coupling in a good range of kinematic space based on the eikonal approximation~\cite{Aoude:2020mlg,Chen:2021huj}.

In this paper we would like to study the eikonal phase in greater detail in light of the quantum information perspective, and inspect a related but orthogonal aspect, namely the scrambling behaviour of gravitational scattering. A quantitative measure of the scrambling power of a unitary evolution is given by the tripartite information~\cite{Hosur:2015ylk}. 
We will pay special attention to expanding the result analytically in $G_N/|b|^2$ particularly for spin-$1/2$ and spin-$1$ scattering particles, where $G_N$ is the Newton's constant, and $b$ is the impact parameter between the scattering.
Surprisingly, we find that bounds of the tripartite information through computation of Renyi entropies show universal behaviours independent of theories describing the gravitational background, in both the non-relativistic and the high-energy limits. We compute a large-spin result that confirms that scrambling tends to diminish in this limit, reducing to a more classical behaviour. 
The tripartite information can be computed analytically as an expansion at least for spin-$1/2$ and spin-$1$. With such analytic control, we are able to identify a term logarithmic in the Newton's constant. This can be contrasted with logarithmic loop corrections to black hole entropies notably discussed in \cite{Sen:2012dw}.
The scrambling power is a minimum at minimal coupling only in the relativistic limit however.  
We obtain also an analytical large-spin limit of the Renyi-2 tripartite entropy. It shows that generically for large impact parameter $b$ and leading order in $G_N$, scrambling power decreases exponentially with large spins. This is perhaps a confirmation of intuition that large spins lead to more classical behaviour. Although for spins much greater than the product of the impact parameter and the particle mass, quantum behaviour would win and scrambling increases exponentially instead.

We also revisit the question studied in~\cite{Aoude:2020mlg,Chen:2021huj}, but consider how entanglement generation is related to the initial states of the scattering particles.
It is customary that we take the spin coherent states as the initial states of semi-classical spinning objects. (See for example \cite{Aoude:2021oqj,Gonzo:2022rfk}.)  
Here, we explore a family of squeezed spin coherent states, and see if the entanglement generation should single out a squeezed state that is most resistant to getting entangled, and thus most classical. This is inspired by considerations in the literature that study squeezed coherent states in harmonic oscillators.  (For a review, see \cite{Ma_2011}.) We again adopt this perturbative approach that allows analytic treatment. We find that entanglement generation is at its minimum or maximum at vanishing squeezing, depending on the relative orientation of the spins of the two particles. While this analysis does not single out the zero-squeezed state, it points to possibilities of amplifying quantum effects by appropriate adjusting of initial states.

Our paper is organised as follows. In section~\ref{sec:RevAmpEikonal}, we review the tree-level amplitude of the $2\rightarrow 2$ scattering process of two spinning blackholes and the corresponding eikonal approach. In section~\ref{sec:scrambling}, we study the scrambling power of this scattering process. We display the bounds for the tripartite information and its large-spin limit. We also compute the exact tripartite information for the spin-$1/2$ case. In section~\ref{sec:coherent}, we consider different initial states and study how classically they behave through the lens of entanglement.

\section{Gravity amplitudes and the Eikonal approximation}\label{sec:RevAmpEikonal}
In this section, we first review the tree-level amplitude for the $2\rightarrow 2$ scattering of a spinning binary system interacting via gravity in the classical (soft) limit. The eikonal phase can be extracted from the amplitude Fourier transformed to impact parameter space, sometimes called the eikonal amplitude. 

In the center-of-mass frame, the momenta of the incoming particles are parameterized by
\begin{align}
p_1 = (E_1, 0, 0, p) \,,  \; p_2 = (E_2, 0,0,-p) \,, 
\end{align}
whereas the outgoing particles are labeled as $p'_1$ and $p'_2$ respectively. 
The total momentum transfer is then $q=(0,\vec{q}) = p'_1-p_1$. In the classical limit $\hbar\rightarrow 0$, we consider only the long-range interaction which is given by the soft graviton exchange, namely $q^2\sim\mathcal O(\hbar^2)$. Hence at tree level, only the $t$-channel with the pole $q^2\rightarrow 0$ contributes to the $2\rightarrow 2$ scattering. Consequently we have $p_1\cdot q= p_2\cdot q=q^2/2 \rightarrow 0$ in the classical limit.

For massive particles of generic spins $s_1$ and $s_2$, the tree-level amplitude is computed in~\cite{Arkani-Hamed:2017jhn,Chung:2019duq,Chung:2020rrz} and expressed in terms of the Wilson coefficients~\cite{Levi:2015msa,Levi:2016ofk},
\begin{align}\label{eq:Moriginal}
M_{\text{tree}}(q^2) = -8\pi  G_N {m_1^2 m_2^2 \over q^2} \sum_{\eta=\pm 1} e^{2\eta\Theta} \left[\varepsilon^*_{1'} W(\eta\tau_1)\varepsilon_1\right] \left[\varepsilon^*_{2'} W(\eta\tau_2)\varepsilon_2\right] +\mathcal O(q^0)
\end{align}
where $\varepsilon_i$ denotes the polarization tensor of the spinning particle $i$ and the sum is taken over the helicities of the graviton exchange. We have adopted the notations in~\cite{Chung:2019duq,Chung:2020rrz,Aoude:2020mlg,Chen:2021huj} where $\tau_i = {q\cdot S_i \over m_i}$ and $\cosh\Theta = {p_1\cdot p_2 \over m_1 m_2}$. Here $S_i$ denotes the Pauli-Lubanskin spin vector with respect to particle $i$ and the function $W$ above is defined as
\begin{align}\label{eq:Woriginal}
W(\eta\tau_i) = \sum_{n=0}^{2 s_i} {C_{i,n} \over n!} \left(\eta {q\cdot S_i \over m_i}\right)^n\,,
\end{align}
where $C_{i,n}$ denotes the Wilson coefficients describing the multi-pole moments of the spinning particles. Rotating black hole solutions correspond to the minimal coupling given by $C_{1,n}=C_{2,n}=1$. We note that the spins here are perceived in the classical sense, which we shall discuss in more detail shortly. 
The spin vector can be transformed to an operator $\mathbb{S}^\mu$~\cite{Maybee:2019jus,Chung:2019duq} 
\begin{align}
\mathbb{S}^\mu_i\equiv \varepsilon_i^* S^\mu_i \varepsilon_i = \left( {\vec{p}_i\cdot \vec{\Sigma}  \over m_i} ,~ \vec{\Sigma}+{\vec{p}_i\cdot \vec{\Sigma}  \over m_i (m_i+E_i)}\vec{p}_i\right)\,,
\end{align}
which acts on the little group indices with $\Sigma$ denoting the spin operator in the rest frame. Re-expressing (\ref{eq:Woriginal}) in terms of $\mathbb S$ requires an extra Thomas-Wigner factor for each particle to compensate the rotation in the little group space. (For detailed discussions, see \cite{Chung:2019duq,Chung:2020rrz}.) The Thomas-Wigner factor reads
\begin{align}
U_i = \exp\left[ -i {m_1 m_2 \mathbb{E}_i \over r_i (E_1+E_2)}\right] \,,
\end{align}
where we have $r_a =1+ E_a / m_a$, $\mathbb{E}_1 \equiv {1 \over m_1^2 m_2} \,\epsilon_{\mu\nu\rho\sigma} q^\mu p_1^\nu p_2^\rho\, \mathbb{S}^\sigma_1$ and $\mathbb{E}_2 \equiv {1 \over m_1 m_2^2} \,\epsilon_{\mu\nu\rho\sigma} q^\mu p_1^\nu p_2^\rho\, \mathbb{S}^\sigma_2$. Consequently, the amplitude (\ref{eq:Moriginal}) is rewritten as follows,
\begin{align}\label{eq:Mbar}
\bar{M}_{\text{tree}}(q^2) = -8\pi G_N {m_1^2 m_2^2 \over q^2 } \sum_{\eta=\pm 1} e^{2\eta\Theta} W_1 (\eta \mathbb{T}_1) W_2 (\eta \mathbb{T}_2) U_1 U_2 +\mathcal O(q^0)\,,
\end{align}
where we have adopted the notation $\mathbb{T}_i = {q\cdot \mathbb S_i \over m_i}$ and the bar denotes the amplitude with the Thomas-Wigner rotations taken into account.

In the classical limit $q^2\rightarrow 0$, the matrices $\mathbb{E}_i$ can be analytically continued~\cite{Chung:2020rrz,Guevara:2017csg}
\begin{align}
\mathbb{E}_i = -i\sinh\Theta\, \mathbb{T}_i\,.
\end{align}
This relation allows us to trade $\mathbb E_i$ and $\mathbb T_i$ in (\ref{eq:Mbar}) and it turns out the most convenient to rewrite all the $\mathbb E_i$ factors in terms of $\mathbb T_i$ for our purposes. In later parts of the paper, we have all implicitly adopted this treatment when referring to (\ref{eq:Mbar}).

Generally speaking, in the classical limit, the so-called classical eikonal phase, denoted as $\chi$ can be extracted from the amplitudes in impact parameter space. The following relation is established in the non-spinning case and it is conceivable that the same structure carries over to the spinning case,
\begin{align}\label{eq:ampvseikonal}
1+ i \mathcal M(b) = \left(1+i\Delta(b) \right) e^{i\chi(b)}\,,
\end{align}
where $\mathcal M$ denotes the Fourier transform of the amplitude and $\Delta(b)$ denotes the quantum remainder~\cite{DiVecchia:2021bdo,Heissenberg:2021tzo,DiVecchia:2022nna}. The quantum remainder enters at one loop and therefore at tree level, the leading order of the classical eikonal phase is simply the Fourier transform of the amplitude
\begin{align}\label{eq:eikonal0}
\chi^{(0)}(b) = {1\over 4 |p| E}\int {d^{2} \vec{q} \over (2\pi)^{2}} e^{i\vec{q}\cdot\vec{b}} \bar{M}_{\text{tree}}(q^2) \,,
\end{align} 
where we note that both $b^\mu$ and $q^\mu$ are space-like vectors in the $xy$-plane only. Hence the Fourier transform is performed in $2$ dimensions. 
Beyond tree level, the relation between the classical eikonal phase and the amplitude in impact parameter space is less explicit, but can be read off from (\ref{eq:ampvseikonal}) order by order. 

As we shall see in later sections, the hierarchy of scales encountered in the scattering process is crucial to the validity of our analyses. Let us elaborate on this by first restoring $\hbar$ explicitly in the quantities we have encountered~\cite{Kosower:2018adc,Chung:2019duq},
\begin{align}
q\rightarrow \hbar q\,,~~~ S^\mu_{1,2} \rightarrow {S^\mu_{1,2} \over \hbar}\,,~~~G_N\rightarrow {G_N \over \hbar}\,,~~~ m_{1,2}\rightarrow m_{1,2}\,,~~~ p_{1,2}^\mu \rightarrow p_{1,2}^\mu\,.
\end{align}
For notational brevity, keeping the above in mind, later in this paper we still work in the unit $\hbar =1$ and the classical limit can be equivalently viewed as a large-impact parameter limit. In other words, $1/|b|$ plays the role of $\hbar$.
Schematically, the amplitude in impact parameter space admits a double expansion in both $G_N$ and $|b|$ as follows, 
\begin{align}
i \mathcal M(b) =  G_N {f^{\text{cl}}_0(b,S) \over |b|^{-2\epsilon}} +G_N^2 \left[ {f^{\text{scl}}_1(b,S) \over |b|^{-4 \epsilon}} + {f^{\text{cl}}_1(b,S) \over |b|^{1-4 \epsilon}} + {f^{\text{q}}_1(b,S) \over |b|^{2-4\epsilon}}\right] + \mathcal O(G_N^3)\,,
\end{align}
where $f$'s are functions of the kinematics and spin degrees of freedom, $S^\mu$ denotes the spin operator of either of the particles and $\epsilon$ denotes the regulator in dimensional regularization. The superscripts indicate the so-called \emph{superclassical}, \emph{classical} and \emph{quantum} contributions.
It is straightforward to observe from the tree-level amplitude~(\ref{eq:Mbar}) that in $f_0^{\text{cl.}}$ for each spin operator acquires a factor of $(\hat{b}\cdot S)/ (|b|m_{1,2})$ where $\hat{b}^\mu = b^\mu/|b|$. This factor is taken to be of $\mathcal O(1)$. 
We have also checked that $f_1^{\text{cl.}}(b,S)$ is a sum over monomials of the same power counting, namely a scalar containing one power of $\hat{b^\mu}/|b|$ and one power of $S^\mu$
~\cite{Bern:2020buy,Bern:2022kto,Chung:2019duq,Chung:2020rrz,Chen:2021qkk}. These monomials are all of $\mathcal O(1)$. Schematically we will denote them all as $(\hat{b}S)/(|b| m_{1,2})$. We believe that the behaviour of the spin-dependence in $f^{\text{scl.}}_1(b,S)$ can be obtained from (\ref{eq:ampvseikonal}) in order to have the consistent exponentiation~\cite{DiVecchia:2021bdo,Heissenberg:2021tzo,DiVecchia:2022nna}.
Consequently, although the phase $\chi$ is only an expansion in $G_N$ and is universal in $\hbar$ or $1/|b|$,  higher-order corrections to the factor $e^{i\chi}$ contain superclassical terms which are exponentiation of lower orders. They are usually discarded when discussing the classical EFT of a quantum field theory, as they provide no new information. However they are present in the eikonal amplitudes nonetheless and are in fact stronger than the classical contributions. In the following sections of this manuscript, we will see that these superclassical terms, together with those arising from $\chi\chi^\dagger$ which are of the same order in both $G_N$ and $1/|b|$, give the leading contribution to entropies induced by the eikonal amplitudes.

In order to discuss the non-relativistic and the high-energy limits of these entropies, we take a closer look at the dependence of $m_{1,2}\sim m$ and $p = |\vec{p}_1| = |\vec{p}_2|$ in the tree and one-loop amplitudes~\cite{Bern:2020buy,Bern:2022kto,Chung:2019duq,Chung:2020rrz,Chen:2021qkk}. In the non-relativistic regime $p\ll m$ ($\cosh\Theta \rightarrow 1$),  we count the powers of $m$ and $p$ in the amplitudes and have
\begin{align}
i\mathcal M(b) = {G_N m^3\over |b|^{-2\epsilon} p} \sum_{n} a_{0,n}^{\text{cl}} x^n + {\color{gray}{G_N^2 m^6\over |b|^{-4\epsilon} p^2} \sum_{n} a^{\text{scl.}}_{1,n} x^n} + {G_N^2 m^4 \over |b|^{1-4\epsilon} p} \sum_{n} a^{\text{cl.}}_{1,n}x^{n} + \cdots\,,
\end{align}
where $x = (\hat{b}S)/ (|b|m_{1,2})\sim\mathcal O(1)$ as defined above. We note that here the scaling behaviour of the one-loop superclassical terms in gray is inferred from the consistency requirement that $f^{\text{scl}}_{1}\propto (f_0^{\text{cl}})^2$. Demanding that the superclassical contribution is indeed stronger than the classical one, we have $m/p \gg 1/(|b|m)$ which is naturally satisfied in the non-relativistic regime.

Similarly in the high-energy (ultra-relativistic) regime $p\gg m$ ($\cosh\Theta\sim p/m$), the amplitudes behave as follows,
\begin{align}
i\mathcal M(b) = {G_N p^2\over |b|^{-2\epsilon}} \sum_{n} a_{0,n}^{\text{cl}} x^n + {\color{gray}{G_N^2 p^4\over |b|^{-4\epsilon}} \sum_{n} a^{\text{scl.}}_{1,n} x^n} + {G_N^2 m p^2 \over |b|^{1-4\epsilon} p} \sum_{n}a^{\text{cl.}}_{1,n} x^{n} + \cdots\,.
\end{align}
The same analysis gives that $(p/m)^2 \gg 1/(|b|m)$ guarantees that the classical contribution would not spoil our later computations.

\section{Scrambling spinning particles via gravitational interactions}\label{sec:scrambling}
    
In this section, we would like to inspect the scrambling power of a perturbative gravitational scattering event. 
As reviewed in the previous section, in the classical limit, the eikonal amplitude $\exp(i \chi )$ approximates the evolution of a pair of particles scattered through soft gravitons. When the particles carry non-trivial angular momenta, the eikonal amplitude is a unitary matrix in spin space. Therefore it is natural to treat $\exp(i \chi )$ as a unitary gate and inspect the scrambling power of such a gate.

There are quite a number of quantitative measures of scrambling power. A particularly handy one is the tri-partite information defined in~\cite{Hosur:2015ylk}. 
The idea is that in the presence of scrambling, input information would be highly delocalised by the evolution, and one needs the entirety of the output qubits to recover the original state before the input qubits. This is a statement about the particular pattern of correlation between output degrees of freedom and input degrees of freedom. A quantitative measure of correlation in a quantum state is ``entanglement entropy''.  We however have a unitary operator here and we would like to measure the correlation pattern between the input and out put indices. Therefore, we consider treating the unitary gate as a wavefunction, and the spin degrees of freedom of the scattering particles as qubits. Then one can utilise the toolbox available to measuring correlation or entanglement in a state. Utilising the defining property of scrambling, one can divide the input qubits made up of the in-coming scattering particles into two sets $A$ and $B$, and the output qubits into two sets $C$ and $D$, and compute the tripartite information $I(A, C, D)$ defined as
\begin{align}
I(A,C,D) & :=  I(A:C) + I(A:D) - I(A:CD) \nonumber \\
&=S_A + S_C + S_D - S_{AC} - S_{AD} - S_{CD} + S_{ACD} ,
\end{align}
where $S_{X}$ is the entanglement entropy of region $X$ with the rest of the system, where $I(X:Y) := S(X) + S(Y) - S(XY)$ is the mutual information between $X$ and $Y$, which measures the information of X stored in Y. The tripartite information thus essentially measures the amount of information in $A$ that is stored non-locally in $C$ and $D$. 
As in, this is the difference between information in A that can be restored in C and D together, and the total amount of information in A that can be individually recovered in C and D separately. 
We note that in the case at hand, $A$ and $B$ would naturally be taken as the spin space of {\it input} particles 1 and 2 respectively, and $C$ and $D$ taken as the respective spin spaces of particles 1 and 2 {\it after} the scattering event.

To compute the entanglement entropies needed above, we need to construct the reduced density matrix. Treating the eikonal as a ``wavefunction'' as prescribed above, we have
\begin{equation}
|\chi\rangle \equiv  \frac{1}{(2s_1 + 1)(2s_2 +1)} \sum_{i_1,i_2, j_1,j_2} \mathcal{E}_{i_1, i_2}^{j_1, j_2} | i_1, i_2, j_1, j_2\rangle, \qquad \mathcal{E}  \equiv \exp(i \chi),
\end{equation}
where we have restored the spin indices explicitly after the first equality, and $i_a , j_a \in \{-s_a, -s_a +1, \cdots, s_a \}$ are the spin indices of particle $a$ before and after scattering respectively. 
The density matrix is thus given by
\begin{equation}
\rho \equiv   \sum_{i_1,i_2,j_1,j_2, i'_1, i'_2, j'_1, j'_2} \mathcal{E}_{i_1,i_2,j_1,j_2}   | i_1, i_2, j_1, j_2\rangle  \langle  i'_1, i'_2, j'_1, j'_2 | \mathcal{E}^*_{i'_1,i'_2,j'_1,j'_2} . 
\end{equation}
Now suppose we would like to compute $S_A$, we obtain the reduced density matrix by tracing out $B,C,D$. 
This gives
\begin{equation}
\rho_A = \frac{1}{(2s_1+1)(2s_2+1)} \sum_{i_1, i'_1}    \mathcal{E}_{i_1, i_2, j_1,j_2} \mathcal{E}^*_{i_1', i_2, j_1, j_2} |i_1\rangle\langle i'_1| = \frac{1}{(2s_1+1)}\sum_{i_1, i_1'} | i_1\rangle \langle i_1|,
\end{equation}
where we have made use of the unitarity of $\mathcal{E}$ as a matrix, so that $\mathcal{E}^*_{i_1', i_2, j_1, j_2} = \mathcal{E}^{-1}_{j_1,j_2, i_1', i_2}$. 
This means that the spin $A$ is maximally entangled with the rest of the spins. In this case the entanglement entropy has no dependence on the kinematics and the couplings, and it is given simply by
\begin{equation}
S_A = \ln(2s_1 + 1). 
\end{equation}
By the same token, $S_B, S_D, S_{ABD} \equiv S_C, S_{AB} \equiv S_{CD}$  all take the form of $S_X = \log(d_X)$, where $d_X $ is the dimension of the reduced spin space $X$. $S_{AD}$ receives no $\mathcal O(G_N^2)$ contributions either due to unitarity.

The only non-trivial contributions therefore come from $S_{AC}$.
As discussed in Section~\ref{sec:RevAmpEikonal}, $\mathcal{E}$ admits a double expansion in both $G_N$ and $|b|$, and consequently we work in similar double expansions throughout this paper. It is straightforward to check that the terms linear in $G_N$ cancel out in all the reduced density matrices above, when the density matrices are properly normalized. At $\mathcal{O}(G_N^2)$, $\mathcal{E}$ receives the superclassical ($\mathcal O(|b|^{4\epsilon})$) and classical ($\mathcal O(|b|^{-1+4\epsilon})$) contributions. Despite that the superclassical contribution comes purely from the exponentiation of the tree-level eikonal phase $\chi^{(0)}$, it is still the leading term at $\mathcal O(G_N^2)$ due to the classical (large impact parameter) limit, provided that the conditions given at the end of Section~\ref{sec:RevAmpEikonal} are satisfied. Therefore, with the validity of the regime in mind, we ignore the contribution from the one-loop amplitude and consider only the leading term which arises from the tree-level amplitude. As we are going to show, it is possible to obtain exact analytical results in a $G_N$ expansion.

As a quick summary, the tripartite information takes the following form to order $G_N^2$
\begin{align} \label{eq:tripartite_simplified}
I(A,C,D) = \ln[(2s_1+1)(2s_2+1)] - S_{AC}.
\end{align}
The larger is $S_{AC}$, the smaller is $I(A,C,D)$, which signifies that information is more scrambled.

\subsection{An upper bound}

It is quite cumbersome to compute the entanglement entropy, while one can more readily recover analytic results for Renyi-$r$ entropy $S^{(r)}_X$ for a subsystem $X$, defined as 
\begin{align} \label{eq:renyi}
S^{(r)}_X = \frac{\ln [\textrm{tr}(\rho_X^r)/(\textrm{tr}(\rho_X))^r]}{1-r}.
\end{align}
Note that in the limit $r\to 1$ the Renyi-$r$ entropy recovers the entanglement entropy. They also satisfy a well known inequality
\begin{align}
S^{(1)} \geqslant S^{(2)} \geq S^{(3)} \geqslant \cdots S^{(\infty)}.
\end{align}
Therefore, we can give an upper bound to the tripartite information (\ref{eq:tripartite_simplified})
\begin{align}
I(A,C,D) \leqslant \ln[(2s_1+1)(2s_2+1)]  - S^{(2)}_{AC}.
\end{align}

In principle the Renyi-$r$ entropy $S_{AC}^{(r)}$ can be computed straightforwardly, given the wavefunction $\mathcal E_{i_1 i_2}^{j_1 j_2}$. Here we sketch out the derivation. 
Schematically, the wavefunction $\mathcal E_{i_1 i_2}^{j_1 j_2}$ takes the expansion below
\begin{align}
\mathcal E_{i_1 i_2}^{j_1 j_2} = \; & \delta^{j_1}_{i_1} \delta^{j_2}_{i_2} + G_N \sum_{n_1 n_2} \sum_{I_1 I_2} \mathbb{A}_{n_1 n_2}^{I_1 I_2} \left(O_{I_1}^{(n_1)}\right)_{i_1}^{j_1}\otimes \left(O_{I_2}^{(n_2)}\right)_{i_2}^{j_2} \nonumber\\
& + G_N^2 \sum_{n_1 m_1 n_2 m_2} \sum_{I_1 J_1 I_2 J_2} \mathbb{B}^{I_1 J_1 I_2 J_2}_{n_1 m_1 n_2 m_2} \left( O^{(n_1 m_1)}_{I_1 J_1}\right)_{i_1}^{j_1} \otimes \left( O^{(n_2 m_2)}_{I_2 J_2}\right)_{i_2}^{j_2} + \mathcal O(G_N^3)\,,
\end{align}
where we have packaged the operators acting on the spin states as $O_0 = \mathbb I$, $O_{+,-} = \Sigma_{+,-} $ and the extra indices $i_a, j_a$ denote the incoming and outgoing states. We have introduced the notations:  $O_{I}^{(n)} = \left(O_{I}\right)^n,~O_{IJ}^{(nm)} = \left(O_I\right)^n \left(O_J\right)^m$. The summations are taken over $n_a, m_a = 1,\cdots, 2s_a$, for spin-$s_a$ and the operator summations labelled by $I$'s exclude $\Sigma_z$ , since $\Sigma_z$ does not enter $\chi^{(0)}(b)$, which gives the terms linear in $G_N$ above. This can be observed immediately from the definition of the tree level eikonal phase~(\ref{eq:eikonal0}). Moreover, mixed terms $\Sigma_+^n \Sigma_-^m$ cancel out at $\mathcal O(G_N)$. At $\mathcal O(G_N^2)$, $\Sigma_z$ contributes but only through $(\Sigma_+)^n (\Sigma_-)^m$ for positive $n,m$. As we will see shortly, the contribution linear in $G_N$ drops out in the Renyi-$r$ entropy and therefore we keep the expansion up to the quadratic order. The coefficients $\mathbb A_{n_1 n_2}^{I_1 I_2}$ and $\mathbb B_{n_1 m_1 n_2 m_2}^{I_1 J_1 I_2 J_2}$ are functions of kinematic variables and independent of the spin-related degrees of freedom. They can be read off from the eikonal $\mathcal E$. Before diving into details, we note the immediate simplifications following from the above observations:   
$\mathbb A^{00}_{nm} = 0$ for any $n,m>1$, $\mathbb A^{0I}_{n m} = \mathbb A^{I0}_{mn}=0$ for $I=\pm$ and $n>1$, and $\mathbb A^{IJ}_{nm}=0$ if $\{I,J\}=\pm$ and $I\neq J$. We also note that $\mathbb A^{00}_{11}$ is multiplied by identity operators acting on both spins, the same as the leading order. For our purpose, this term can be absorbed into the leading term and in principle can affect the normalization. Since we keep up to $\mathcal O(G_N^2)$ only, this term does not contribute at all and therefore can be discarded from now on. The quadratic terms correspond to the superclassical contributions and arise from the exponentiation of the tree-level eikonal phase. Hence the coefficients are given by
\begin{align}
\mathbb B_{n_1 m_1 n_2 m_2}^{I_1 J_1 I_2 J_2} = {1\over 2} \mathbb A_{n_1 n_2}^{I_1 I_2} \mathbb A_{m_1 m_2}^{J_1 J_2}\,.
\end{align}

Now the reduced density matrix $\rho_{AC}$ reads
\begin{align} \label{eq:rho}
\rho_{AC} =&  \sum_{i_1, j_1, i'_1, j'_1} \mathcal E_{i_1 i_2 j_1 j_2} \mathcal E^*_{i'_1,i_2, j'_1, j_2 } | i_1 j_1\rangle \langle i'_1 j'_1| \nonumber \\
=& \left(\mathbb{P}  +G_N \Omega_1 +G_N^2 \big(\Omega_2^{(1)}+\Omega_2^{(2)}\big) \right)_{i_1 j_1 i'_1 j'_1} | i_1 j_1\rangle \langle i'_1 j'_1| \,,
\end{align}
where the leading term is a projector $\mathbb P_{i_1 j_1 i'_1 j'_1} = (2s_2+1) \delta_{i_1 j_1} \delta_{i'_1 j'_1} $ and the other operators read
\begin{align}
(\Omega_1)_{i_1 j_1 i'_1 j'_1}   =& \sum_{ \{n_a ,I_a\} }  \left[\mathbb{A}_{n_1 n_2}^{I_1 I_2} \big(O^{(n_1)}_{I_1}\big)_{i_1 j_1} \delta_{i'_1 j'_1} + \big(\mathbb{A}_{n_1 n_2}^{I_1 I_2}\big)^*  \big(O^{(n_1)}_{I_1}\big)_{i'_1 j'_1} \delta_{i_1 j_1}\right] \text{tr}\big(O_{I_2}^{(n_2)}\big)  \,,\\
(\Omega_2^{(1)})_{i_1 j_1 i'_1 j'_1}   = &  \sum_{ \{n_a,I_a\} } \left[ \mathbb{B}_{n_1 m_1 n_2 m_2}^{I_1 J_1 I_2 J_2} \big( O_{I_1 J_1}^{(n_1 m_1)}\big)_{i_1 j_1} \delta_{i'_1 j'_1} + \big(\mathbb{B}_{n_1 m_2 n_2 m_2}^{I_1 J_1 I_2 J_2}\big)^* \big( O_{I_1 J_1}^{(n_1 m_1)}\big)_{i'_1 j'_1}\delta_{i_1 j_1} \right]\nonumber\\
& \quad\quad \times  \text{tr}\big( O_{I_2 J_2}^{(n_2 m_2)} \big)  \,,\\
(\Omega_2^{(2)})_{i_1 j_1 i'_1 j'_1}   = & \sum_{ \{n_1,I_a\} }  \mathbb{A}_{n_1 n_2}^{I_1 I_2}\big( \mathbb{A}_{m_1 m_2}^{J_1 J_2}\big)^*    \text{tr}\big( O_{I_2 J_2}^{(n_2 m_2)} \big) \big( O_{I_1}^{(n_1 )}\big)_{i_1 j_1} \big( O_{J_1}^{(m_1 )}\big)_{i'_1 j'_1}  \,,
\end{align}
where $I_a,J_a \in \{ \mathbb I, \Sigma^\pm_a\}$ and small letters denote the power of these operators $\{ n_a, m_a\}  =1,\cdots, 2s_a$.
Recalling that $\text{tr}\big(\mathbb{I}\big) = 2s_2+1$ when we trace out $B$ and $D$ corresponding to particle $2$ and that $\text{tr}\big(\Sigma^n_\pm\big) = 0$, the remaining traces above are readily computed. In addition to the identity matrix, the only non-vanishing trace is given by
\begin{align}
\text{tr}_2\left( O_{\pm \mp}^{(mm)} \right) = \text{tr}_2 \left( \Sigma^m_+ \Sigma^m_-\right) \equiv t^{(m)}_2
\end{align} 
where the additional subscript indicates that the partial trace is taking over indices belonging to particle 2.

Taking these traces can be carried out straightforwardly and the detailed computation is demonstrated in Appendix~\ref{app.renyi}. Here we present the final expression for the Renyi-$r$ entropy
\begin{align}\label{eq:rRenyi}
S^{(r)}_{AC} = {G_N^2 r\over r-1} {1 \over (2s_1+1) (2s_2+1)} \sum_{m,n} \sum_{I,J=\pm} \mathbb A_{mn}^{IJ} \left( \mathbb A_{mn}^{IJ}\right)^* t^{(m)}_1 t^{(n)}_2\,,
\end{align} 
where $t^{(m)}_a = \text{tr}\, \Sigma_+^m \Sigma_-^m$ with $m=1,\cdots, 2s_a$. Notice that $\mathbb B$-terms, namely the superclassical terms in $\mathcal E$, drop out in the Renyi-$r$ entropy when properly normalised. The $G_N^2$ contribution above come from $\chi^{(0)} \big( \chi^{(0)}\big)^*$ and although at the same order in the soft expansion with these superclassical terms, are of a slightly different nature.

For $s_1 = s_2 = 1/2$, only the minimal coupling interactions are involved and the Renyi-2 entropy $S_{AC}^{(2)}$ reads
\begin{align} \label{eq:spinhalfrenyi}
S_{AC,1/2}^{(2)} &= {G_N^2 \over 16} \left| \mathbb A_{1,1}^{++}\right|^2 =  {16 G^2_N \pi^4 m_1^2 m_2^2 \over (-b^2) E^4 r_1^2 r_2^2}\times \\
& \left[  8E\cosh\Theta \sinh^2 \Theta(m_1 r_2 +m_2 r_2) -4\cosh 2\Theta\,\left(E^2 r_1 r_2 +m_1 m_2 \sinh^2 \Theta\right) \right]^2\,,\nonumber
\end{align}

Similarly, for $s_1=s_2=1$, plugging the values of the traces $t_a^{(m)}$, (\ref{eq:rRenyi}) simply reads
\begin{align}
S^{(2)}_{AC,1} = \frac{64}{9}G^2_N(|\mathbb A^{++}_{1,1}|^2+|\mathbb A^{++}_{1,2}|^2+|\mathbb A^{++}_{2,1}|^2+|\mathbb A^{++}_{2,2}|^2)\,.
\end{align}
We give the explicit expressions for the coefficients $\mathbb A_{mn}^{IJ}$ as follows,
\begin{align}
\mathbb A^{++}_{1,1} &=-\mathbb A^{--}_{1,1}=i\frac{4\pi^2 }{b^2}(\frac{m_{1} m_{2}c_{2 \Theta} s_{\Theta} }{E^{2} r_{1} r_{2}}+\frac{c_{2 \Theta}}{s_{\Theta}}-\frac{2 m_{2} c_{\Theta} s_{\Theta}}{E r_{1}}-\frac{2 m_{1} c_{\Theta} s_{\Theta}}{E r_{2}}) \,,\nonumber  \\
\mathbb A^{++}_{2,1} &=\mathbb A^{--}_{2,1}=\frac{4\pi^2 }{b^3}\left(-\frac{C_{1,2} c_{2 \Theta }}{E r_2}+\frac{2 C_{1,2} c_{\Theta }}{m_1}-\frac{m_2^2 c_{2 \Theta } s_{\Theta }^2}{E^3 r_1^2 r_2}+\frac{4 m_2 c_{\Theta } s_{\Theta }^2}{E^2 r_1 r_2}-\frac{2 m_2 c_{2 \Theta }}{E m_1 r_1}+\frac{m_2^2 s_{\Theta } s_{2 \Theta }}{E^2 m_1 r_1^2}\right)\,, \nonumber\\
\mathbb A^{++}_{1,2} &=\mathbb A^{++}_{1,2}=\frac{4\pi^2 }{b^3}\left(-\frac{C_{2,2} c_{2 \Theta }}{E r_1}+\frac{2 C_{2,2} c_{\Theta }}{m_2}-\frac{m_1^2 c_{2 \Theta } s_{\Theta }^2}{E^3 r_1 r_2^2}+\frac{4 m_1 c_{\Theta } s_{\Theta }^2}{E^2 r_1 r_2}-\frac{2 m_1 c_{2 \Theta }}{E m_2 r_2}+\frac{m_1^2 s_{\Theta } s_{2 \Theta }}{E^2 m_2 r_2^2}\right)\,,\nonumber\\
\mathbb A^{++}_{2,2} &=-\mathbb A^{--}_{2,2}=i\frac{6\pi^2 }{b^4}\left(-\frac{m_2 C_{2,2} c_{2 \Theta } s_{\Theta }}{E^2 m_1 r_1^2}-\frac{m_1 C_{1,2} c_{2 \Theta } s_{\Theta }}{E^2 m_2 r_2^2}-\frac{C_{1,2} C_{2,2} c_{2 \Theta }}{m_1 m_2 s_{\Theta }}+\frac{2 C_{1,2} s_{2 \Theta }}{E m_2 r_2} \right.\nonumber\\
&\left.
+\frac{2 C_{2,2} s_{2 \Theta }}{E m_1 r_1}-\frac{m_1 m_2 c_{2 \Theta } s_{\Theta }^3}{E^4 r_1^2 r_2^2}+\frac{4 m_2 c_{\Theta } s_{\Theta }^3}{E^3 r_1^2 r_2}+\frac{4 m_1 c_{\Theta } s_{\Theta }^3}{E^3 r_1 r_2^2}-\frac{4 c_{2 \Theta } s_{\Theta }}{E^2 r_1 r_2}
\right)\,,
\end{align}
where we have adopted the shorthand notations $c_\Theta=\cosh\Theta$ and $s_\Theta=\sinh\Theta$.

There are several limits one can take. The most immediate ones are the non-relativistic limit $\cosh\Theta \rightarrow 1$ and the opposite, namely the high-energy limit $\cosh\Theta \gg 1$.

\begin{figure}[tbp]
  \centering
  \includegraphics[width=.7\linewidth]{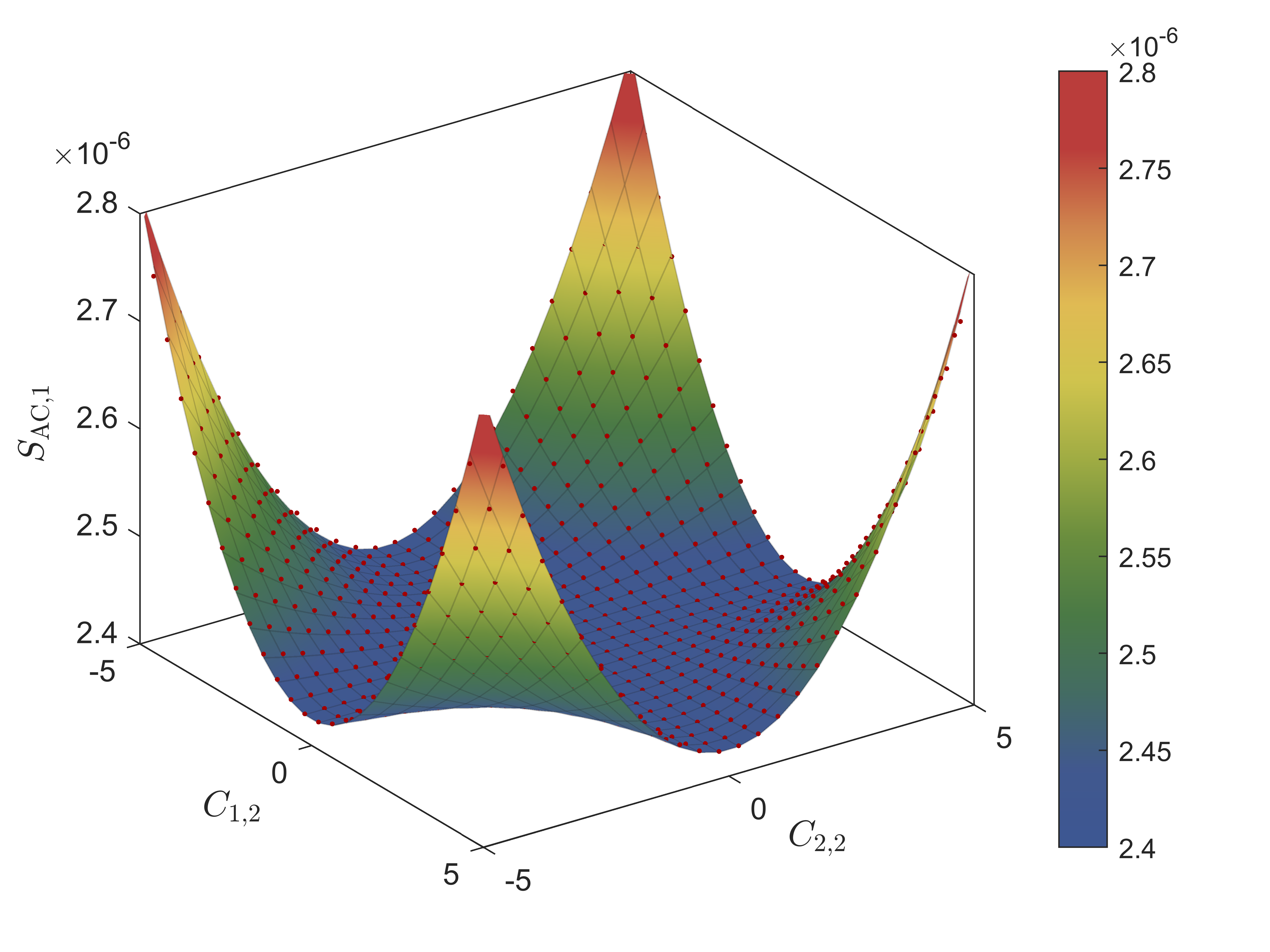}  

\caption{$S_{\text{AC},1}$(red points) and fitted surface vs. $C_{1,2}$,$C_{2,2}$ with $m_1=m_2=m$, $G_N m^2=10^{-5}$, $|b|m=10$, $p/m=0.01$. In this non-relativistic regime, the tripartite information is insensitive to the theory. When the $C_{i,2}$ is comparable to $|b|m$ at the corner, the entropy grows about $20\%$. The surface is fitted by $6.3\times10^{-10}C_{1,2}^2C_{2,2}^2+2.4\times10^{-6}$, which suggests (\ref{eq:non-rel_AC}) gives the right form of the $|b|m$ suppressed theory-dependent sub-leading term. The minimal occurs at $(0,0)$.}
  \label{fig:smallpTI}
\end{figure}

Recall that in the non-relativistic regime $\cosh\Theta\rightarrow 1$, we need to have $m_{1,2}/p \gg 1/(|b| m_{1,2})$ as well. Hence, the Renyi entropies for spin-$1/2$ and spin-$1$ read 
\begin{align} 
&\left. S_{AC, 1/2}^{(2)}\right|_{\text{non-rel}} = {16G_N^2   \pi^4  m_1^2 m_2^2 \over |b|^4 (m_1+m_2)^2 p^2} \,,  \\
&\label{eq:non-rel_AC}\left. S_{AC, 1}^{(2)}\right|_{\text{non-rel}} ={256{G_N^2 \pi^4 m_1^2 m_2^2 }\over {9|b|^4 (m_1+m_2)^2 p^2}}+{\color{BrickRed} \frac{256 G_N^2\pi^4C_{1,2}^2C_{2,2}^2}{|b|^8(m_1+m_2)^2p^2}}\,.
\end{align}
Here we note that in addition to the two inequalities above, this expansion only makes sense when ${|b| m_{1,2} } \ll {m_{1,2} /p}$ is also satisfied. This extra inequality guarantees that although the $G_N^2$ contribution comes at $m_{1,2}^2/p^2$, it is still suppressed by the large impact parameter and the weak coupling constant such that it would not overwhelm the $\mathcal O(G_N^0)$ term before taking the logarithm in (\ref{eq:renyi}).
For spin-$1/2$, we do not have unknown Wilson coefficients; while for spin-$1$, we have two Wilson coefficients $C_{a,2}, a=1,2$ that are theory-dependent. We note that these two coefficients play no role in the non-relativistic limit as they only feature in the first sub-leading term, colored red above, which is suppressed by the large impact parameter.  
Therefore the non-relativistic behaviour is theory-agnostic to leading order.  Note that the Renyi entropy increases with decreasing $p/m_{1,2}$, suggesting that the scrambling power increases with larger masses in the non-relativistic limit.  Figure \ref{fig:smallpTI} is a plot of the entanglement entropy $S_{\textrm{AC}}$ in this regime, which shows the same qualitative behaviour.  

The high-energy regime  $\cosh\Theta\gg 1$ (namely $p\gg m_{1,2}$), in which we also need to demand $(p/m_{1,2})^2\gg 1/(|b|m_{1,2})$, is trickier and we separate our discussions into different scenarios. 

\paragraph{Scenario I} The relevant scales satisfy the following inequalities: $(p/m_{1,2})^2\gg 1/(|b|m_{1,2})$, $\cosh\Theta\gg 1$, and $ p/m_{1,2} \ll |b| m_{1,2}$. Within the validity of these inequalities, we find that the Renyi-2 entropies behave as follows for spin-$1/2$ and spin-$1$, 
\begin{align}
\!\!\!\!\left. S_{AC, 1/2}^{(2)}\right|_{\text{high-energy}}  =&  {256 G_N^2 \pi^4\over |b|^4} - {256 G_N^2 \pi^4 (m_1+m_2)\over |b|^4 p} \,,\\
\!\!\!\!\left. S_{AC, 1}^{(2)}\right|_{\text{high-energy}}  = & {16384 G_N^2 \pi^4 \over  9|b|^4} -{16384 G_N^2\pi^4 (m_1+m_2) \over  9|b|^4 p}\,.
\end{align}

For both spin-$1/2$ and spin-$1$, the first sub-leading term is suppressed by $m_{1,2}/p \ll1$. For spin-$1$, the Wilson coefficients enter through the second sub-leading term which we have not shown. 
We note that the leading behaviour of the Renyi-2 entropy is theory independent and the scrambling power increases with increasing energy. Moreover, the first sub-leading behaviour is still theory-agnostic. Theory dependence only comes at a sub-sub-leading term. 

\begin{figure}[tbp]
  \centering
  \includegraphics[width=.7\linewidth]{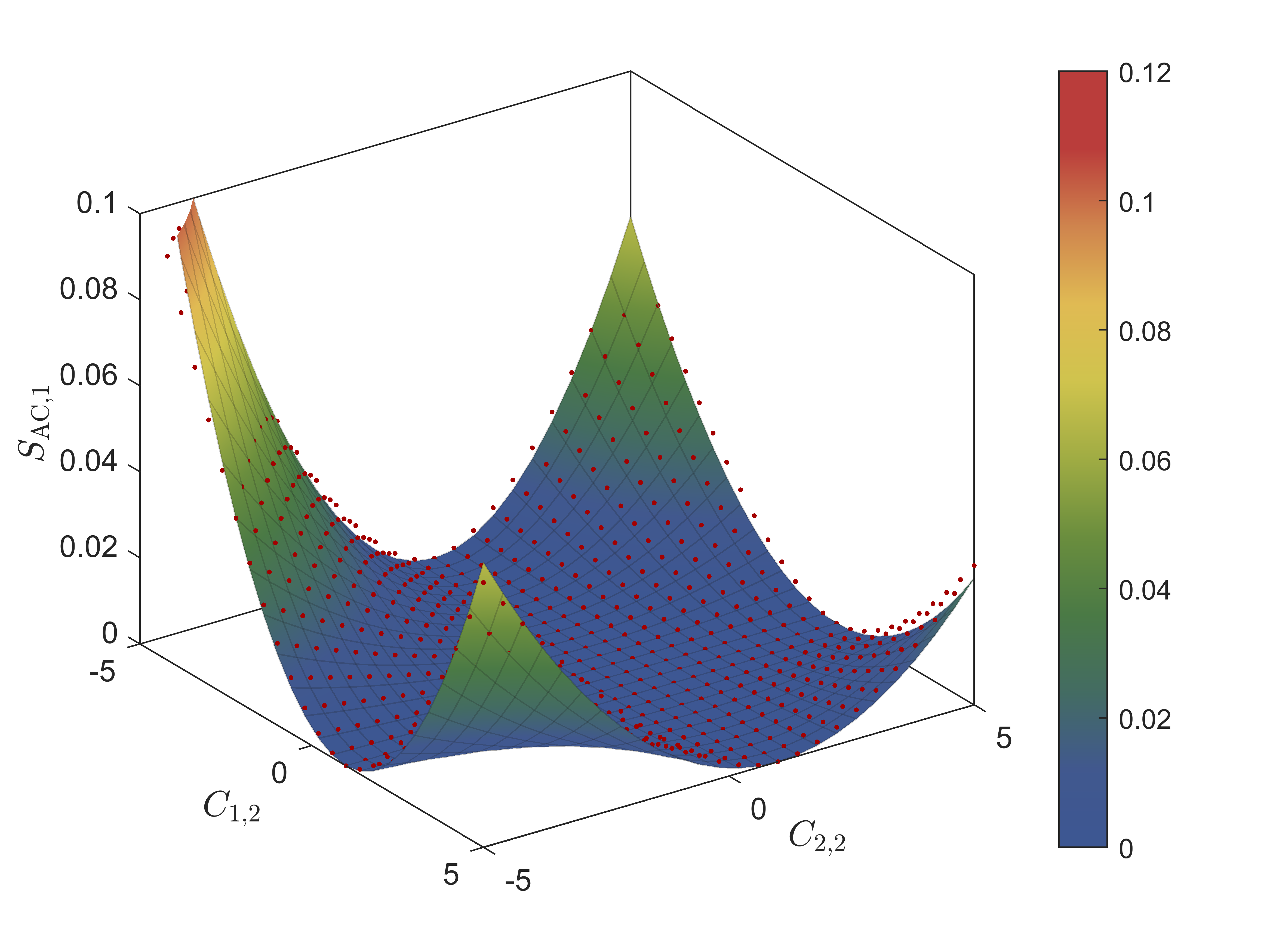}  
 \caption{$S_{\text{AC},1}$(red points) and fitted surface vs. $C_{1,2}$,$C_{2,2}$ with $m_1=m_2=m$, $G_N m^2=10^{-5}$, $|b|m=10$, $p/m=100$. In this Scenario II of the high-energy regime, the tripartite information is sensitive to the theory. The surface is fitted by $0.000115(C_{1,2}-1)^2(C_{2,2}-1)^2$, which suggests that (\ref{eq:SACrel2}) gives the right form of the leading term. The minimal occurs at $(1,1)$.}
 \label{fig:largepTI}
\end{figure}

\paragraph{Scenario II}  The relevant scales satisfy the following inequalities: $(p/m_{1,2})^2\gg 1/(|b|m_{1,2})$, $\cosh\Theta\gg 1$, and $ p/m_{1,2} \gg |b| m_{1,2}$. The expansion of the Renyi-2 entropies then reads,
\begin{align} 
\!\!\!\!\left. S_{AC, 1/2}^{(2)}\right|_{\text{high-energy}} =& {256 G_N^2 \pi^4\over |b|^4} - {256 G_N^2 \pi^4 (m_1+m_2)\over |b|^4 p} \,,\\
\!\!\!\!\label{eq:SACrel2}\left. S_{AC, 1}^{(2)}\right|_{\text{high-energy}}  = & {\frac{4096 G_N^2\pi^4p^4}{b^8m_1^4m_2^4}(C_{1,2}-1)^2(C_{2,2}-1)^2}\nonumber\\
&+{16384 G_N^2 \pi^4  p^2 \left(m_2^4(C_{1,2}-1)^2 + m_1^4(C_{2,2}-1)^2 \right) \over 9 |b|^6 m_1^4 m_2^4}\,.
\end{align}
The expansion takes the same form as in the previous scenario for spin-$1/2$ due to the absence of theory-dependent Wilson coefficients. For spin-$1$, the leading and sub-leading contributions are both theory-dependent and the scrambling has a minimum at the minimal coupling. Figure \ref{fig:largepTI} is a plot of the entanglement entropy in this regime.

In all cases we note that the scrambling ability always decrease with increasing impact parameter, as expected. When there is hardly any interaction between particles, they cannot be possibly scrambled. 

\subsection{Large-spin expansion}
Here we discuss the Renyi entropies and the resulting tripartite information for the minimal coupling in the large-spin limit.

The kinematics-dependent coefficients $\mathbb A_{mn}^{IJ}$ can be read off from the definition given in~(\ref{eq:Woriginal})-(\ref{eq:eikonal0}). The remaining ingredient is the trace $t^{(m)}$. Recall that for a given spin-$s$ and generic $m$, we have $\Sigma_\pm |s,m\rangle = \sqrt{s(s+1) - m(m\pm1)} |s,m\pm 1\rangle$ and hence  
\begin{align}\label{eq:trpmDef}
t^{(m)} \equiv \sum_{n=-s}^{s} \langle s,n| \Sigma_+^m \Sigma_-^m |s,n\rangle = \sum_{n=-(s-m)}^{s} {\Gamma( s+n+1) \Gamma(s+m-n+1) \over \Gamma(s-m+n+1) \Gamma(s-n+1)}\,,
\end{align}
where in the second equal sign we have rewritten products of consecutive integers using the gamma functions. Carrying out the sum, we obtain\footnote{Summing up the summand~(\ref{eq:trpmDef}) in terms of Gamma functions in \textsf{Mathematica} leads to a missing overall ${1\over 2}$ which we have fixed numerically.}
\begin{align}\label{eq:trpm}
t^{(m)}= {-\sqrt{\pi}\, \csc(2s\pi)\Gamma(-m-{1\over2}) \Gamma(m+1) \over 2^{2m+1}\,\Gamma(-2s-m-1)\Gamma(2s-m+1) }={\sqrt{\pi}\,\Gamma(m+1)\Gamma(2s+m+2) \over 2^{2m+1}\, \Gamma(2s-m+1) \Gamma(m+{3\over 2})}\,.
\end{align}
where in the second equal sign we have used $\Gamma(x) \Gamma(1-x) = \pi/\sin(\pi(x+1))$ to remove the singularities of the gamma functions. 

In the large-spin limit $s\rightarrow \infty$, we note that $\Gamma(m+1)/\Gamma(m+3/2)] \sim m^{-1/2}$ in the large $m$ limit. At the same time $\Gamma(2s+m+2)/\Gamma(2s-m+1)$ is increasing with $m$ exponentially quickly. Therefore we conclude that in the large-$s$ limit, the $m=2s$ term should dominate exponentially. That is
\begin{align}
t^{(2s)} = {\pi} (2s/e)^{4s} [1+ \mathcal{O}(1/s)].
\end{align}
Hence, we only need to consider the dominant contribution to the Renyi-$r$ entropies $S^{(r)}_{AC}$ and $S^{(r)}_{AD}$, namely the $m=n=2s$ term in summation in the large-spin limit
\begin{align}
\left. S^{(r)}_{AC} \right|_{s\rightarrow \infty} = {r\over r-1}{G_N^2 \over (2s+1)^2} \left(t^{(2s)}\right)^2 \sum_{I,J=0}^2 \mathbb A_{2s,2s}^{IJ} \left( \mathbb A_{2s, 2s}^{IJ}\right)^*
\end{align}

In the case of minimal coupling, the coefficients $\mathbb A_{2s,2s}^{IJ}$ can be readily extracted. It is given by
\begin{align}
\mathbb A_{2s,2s}^{II} &=  -\frac{32\pi^2  (4s-1)!}{m_a^{2s-2}m_b^{2s-2}( (2s) !)^2  b_I^{4s}}  \left[e^{\Theta}(1- X_a)^{2s}(1-X_b)^{2s} + e^{-\Theta}(1+ X_a)^{2s}(1+ X_b)^{2s} \right] \nonumber\\
&\sim  -\frac{8\pi^2  2^{4s}}{ \sqrt{2\pi s^3}m_a^{2s-2}m_b^{2s-2}  b_I^{4s}}  \left[e^{\Theta}(1- X_a)^{2s}(1-X_b)^{2s} + e^{-\Theta}(1+ X_a)^{2s}(1+ X_b)^{2s} \right]\,,\nonumber\\
\end{align}
where $X_a =  - m_a m_b \sinh\Theta/(E(m_a + E_a)) $ and $I=\pm$. Here we have used $b_\pm = b_1 \pm i b_2$.

Altogether, this gives a large-$s$ behaviour of 
\begin{align}
S^{(2)}_{AC} \sim \frac{G_N^2 \pi^5   (4s/e)^{8s}    \left[e^{\Theta}(1- X_a)^{2s}(1-X_b)^{2s} + e^{-\Theta}(1+ X_a)^{2s}(1+ X_b)^{2s} \right]^2}{2s^5 m_a^{4s-4}m_b^{4s-4}} \left(\frac{1}{b_+^{8s}} + \frac{1}{b^{8s}_-}\right).
\end{align}
We note that the result almost completely depends on the magnitude of $\sigma:= s/(|b| \sqrt{m_a m_b})$ which is raised to a power of $8s$. If it is larger than 1, $S^{(2)}_{AC}$ would grow rapidly with spins, and if it is less than 1, it would be suppressed exponentially by the spin. 
This suggests an interesting competition between the number of degrees of freedom available for scrambling and the strength of the gravitational interaction weakening with increasing separation. 
When gravity is sufficiently suppressed, larger spins scrambling is further suppressed by spin. This confirms with folklore that systems with large quantum numbers should resemble classical ones. Recall that in many quantum mechanical systems when quantum numbers get large they begin to appear continuous again, and that their variance is also suppressed, thus resembling classical systems.  On the other hand, when there are sufficiently large number of degrees of freedom supplied by the spin degrees of freedom such that $\sigma >1$, it can actually exhibit more scrambling, and thus {\it more quantum behaviour}, defying folklore. The issue is that entanglement entropy is restricted by the dimension of the Hilbert space. The computation suggests that there is a window that enhances scrambling, as long as there is a large Hilbert space with sufficiently strong gravitational interaction (i.e. $|b|$ does not go to infinity faster than the spin $s$) .

\subsection{An exact $G_N^2$ result at spin-$1/2$ and spin-$1$}

When the spins of the particles are not too large, it is not difficult to recover the leading $G_N$ contribution to the {\it exact} entanglement entropy $S_{AC}$, which leads to exact results for the tripartite information. For illustrative purpose,
we present the result where the spins of the scatterinng particles are both spin 1/2 and both spin 1.
 
To compute the entanglement entropy, we first solve for the eigenvalues of the reduced density matrix (\ref{eq:rho}) as a $G_N$ expansion. 
When both particles are spin $1/2$, the eigenvalues are listed as follows:
\begin{align}
\lambda_1 =  1-\frac{1}{2}G_N^2|\mathbb A^{++}_{1,1}|^2, \,\, \lambda_2 =  \frac{1}{4}G_N^2|\mathbb A^{++}_{1,1}|^2, \,\, \lambda_3 =\frac{1}{4}G_N^2|\mathbb A^{++}_{1,1}|^2, \,\, \lambda_4 = \mathcal{O}(G_N^3).
\end{align} 

The Renyi - information is then given by
\begin{equation}
    S_{AC,1/2}=\frac{1}{2}G_N^2|\mathbb A^{++}_{1,1}|^2[1+\ln{4} - 2\ln{(G_N|\mathbb A^{++}_{1,1}|)}]
\end{equation}
We note that the $G_N^2$ correction to the largest eigenvalue $\lambda_1$ is equal to half of the Renyi-$2$ entropy.  
This is expected since other smaller eigenvalues of order $G_N^2$ cannot contribute at $G_N^2$ to the Renyi entropy. 
In addition to the $G_N^2$ contribution analogous to
the Renyi entropy computation, we have a logarithmic correction.  The logarithmic correction follows from the fact that $\rho_{AC}$ is not a full rank matrix to zeroth order in $G_N$ (in fact a rank 1 matrix for any spins), which has already been pointed out earlier. 

Using (\ref{eq:spinhalfrenyi}), the logarithmic term takes the form of $+  1/8 G_N^2/(|b|^2 \mathcal{F})^2 \ln[|b|^2/G_N \times \mathcal{F}]$,
where $\mathcal{F}$ denotes dimensionless factors depending on the center of mass energy and the masses of the particles. This can be contrasted with the one-loop correction to the black hole entropy computed for example in \cite{Sen:2012dw}, which contains logarithmic correction of the form $\ln a$, where $a$ is the horizon area measured in Planck units.

This computation can be done for the case of spin-1 particles. One would expect that the logarithmic correction should persist. 
For spin-$1$, the eigenvalues of $\rho_{AC}$ are given by
\begin{align} 
&\lambda_1= 1-\Delta_1, \,\, \lambda_2 = \Delta_1(1+\Delta_2) /4, \,\, \lambda_3 = \Delta_1(1+\Delta_2) /4 , \nonumber \\
&\lambda_4 = \Delta_1(1-\Delta_2) /4, \,\, \lambda_5  = \Delta_1(1-\Delta_2) /4,\,\, \lambda_{i>5} = \mathcal{O}(G_N^3)
\end{align}
where
\begin{align}
\Delta_1&=\frac{32}{9}(|\mathbb A^{++}_{1,1}|^2+|\mathbb A^{++}_{1,2}|^2+|\mathbb A^{++}_{2,1}|^2+|\mathbb A^{++}_{2,2}|^2)G_N^2 \\
\Delta_2&=\sqrt{1-4(\frac{|A^{++}_{1,1}A^{++}_{2,2}|+|A^{++}_{2,1}A^{++}_{1,2}|}{|\mathbb A^{++}_{1,1}|^2+|\mathbb A^{++}_{1,2}|^2+|\mathbb A^{++}_{2,1}|^2+|\mathbb A^{++}_{2,2}|})^2}
\end{align}
Note that $\Delta_1$ is of order $G_N^2$, and $\Delta_2$ is of order $G^0_N$.

This gives $S_{AC}$ for spin-$1$ 
\begin{equation}
    S_{AC,1}=\Delta_1  \left[1+ \ln(\frac{4}{ \sqrt{1-\Delta_2^2}\Delta_1}) - \frac{\Delta_2}{2}\ln(\frac{1+\Delta_2}{1-\Delta_2})\right].
\end{equation}

\section{Semi-classicality and coherent states}\label{sec:coherent}

In the previous sections, we explored how quantum chaotic a gravitational scattering process approximated by the eikonal in the semi-classical limit could potentially be.
The result depends on the properties of the scattering matrix, and made no mention of the initial scattering states.
In the current section, we would like to ask a related but orthogonal question regarding the classicality of the initial states.
In the case of microscopic system, it is possible to prepare the pure initial states. However for macroscopic systems it is more natural to consider mixed states. 
Therefore, we will analyse these two cases separately. In the first part, we will consider a family of squeezed coherent state, and ask if vanisihing squeezing should minimize the final entanglement after scattering. We will also consider a family of mixed states and identify the most classical one.

\subsection{Squeezed spin coherent states}
It is well known that some states behave more classically than others, and they are used as an important ingredient in recovering classical physics. One very important class of examples are coherent states studied extensively in the context of harmonic oscillators. 
The most important characterization of a coherent state for the harmonic oscillator is that it is designed to minimize the uncertainty in the non-commuting variables $\Delta X$ and $\Delta P$ such that 
the Heisenberg uncertainty inequality is saturated. i.e. 
\begin{align}
\Delta X \Delta P = \frac{\hbar}{2}.
\end{align}
In addition to that, the coherent state distributes uncertainty equally between $X$ and $P$ so that
\begin{align}
\Delta X = \Delta P = \sqrt{\frac{\hbar}{2}}.
\end{align}

However, while saturating the Heisenberg uncertainty inequality is certainly a sign of classicality, one could in principle distribute the remaining uncertainty differently between
$X$ and $P$. One can construct a family of states satisfying
\begin{align}
\Delta X = \exp(- \eta) \sqrt{\frac{\hbar}{2}}, \qquad \Delta P = \exp( \eta) \sqrt{\frac{\hbar}{2}},
\end{align}
and they are called the squeezed states. It is a curiosity if any one of these states would be singled out and behave more ``classically''. 
Indeed, the original coherent state corresponding to $\eta =0$ distinguishes itself when one considers an interacting system. 
It is shown in~\cite{Zurek:1992mv} that considering a generic Hamiltonian that couples the harmonic oscillator to a heat bath of harmonic oscillators, the coherent state without squeezing is the most resistant to getting entangled with the bath. 

We would like to ask a similar question here. Is there a spin state that best models macroscopic classical objects carrying angular momentum and interacting via gravity?
In the case of spin states, there is a corresponding notion of {\it spin coherent states}.  For spin $J$, spin coherent states are simply defined as eigenstates of the angular momentum operator $\vec J$ which is aligned in the direction the classical particle is spinning \cite{Giraud_2008, Ma_2011} i.e. for a unit vector $\vec n$,
\begin{align} \label{eq:spin_coherent}
\vert \vec{n}\rangle, \qquad \vec{n}\cdot\vec{J} \vert \vec{n} \rangle =  J  |\vec{n} \rangle\,.
\end{align}
In the case of spins, it satisfies the following uncertainty inequality 
\begin{align}
\Delta J_x  \Delta J_y \leq \frac{ \hbar \langle J_z \rangle}{2}\,.
\end{align}
One can readily check that the above choice of states with $\vec{n} = (0,0,1)$ saturates this inequality, with $\Delta J_x = \Delta J_y = \sqrt{\frac{\hbar \langle J_z \rangle}{2}}$. 

One can also construct squeezed states which distribute uncertainty unevenly between $J_x$ and $J_y$. There are many ways to parametrize squeezed states.
One convenient family is obtained by \cite{Ma_2011}
\begin{align} \label{eq:squeezed_spin}
|\zeta\rangle = \exp(- \zeta J_x^2) | J, J\rangle\,.
\end{align}
This gives $\langle J_x\rangle = \langle J_y\rangle $ =0, preserving the axial symmetry along the $z$- axis. 
We are working in the center of mass frame of the scattering particles, where the in-coming particles have momenta along the $z$-axis. The scattering matrix has axial symmetry along the $z$-direction. We will thus pick an initial state of the form (\ref{eq:squeezed_spin}) for each of the in-coming particles. 
We would like to explore the optimal amount of squeezing that minimizes the entanglement after the scattering process.  
Note that the spin coherent states approaches the coherent state in a harmonic oscillator in appropriate large spin $j$ limits \cite{Ma_2011}, although this is beyond the scope of the current analysis, and could be left for future investigation. 
We will present both analytic and numerical results.

To illustrate this analysis, we take spin-$1$ particles as an example. We would like to consider an initial state that is a direct product state of the two in-coming scattering particles. Moreover, the problem has axial symmetry about the $z$-axis. Therefore we consider states where $J_x$ and $J_y$ have vanishing expectation value. 
A natural family of states would be given by $|\zeta_1\rangle \otimes  |\zeta_2 \rangle$, i.e. in principle the two particles could have different squeezing and they could lead to differing amounts of final entanglement. We would like to look for universal statements that have a cleaner physical interpretation. To that end, we consider small deviations from the spin coherent state where $\zeta$ is vanishing. In the small squeezing limit, we might as well introduce squeezing only to one of the particles.  
Therefore, we would like to consider the two simple initial states below,
\begin{align}\label{def:initState}
|\Psi_1(\zeta)\rangle &=  {e^{-2\zeta}+1\over 2} |+1\rangle\otimes |+1\rangle + {e^{-2\zeta}-1\over 2} |-1\rangle\otimes | +1\rangle\,,\\
|\Psi_2(\zeta)\rangle &=  {e^{-2\zeta}+1\over 2} |+1\rangle\otimes |-1\rangle + {e^{-2\zeta}-1\over 2} |-1\rangle\otimes | -1\rangle\,.
\end{align}
At $\zeta=0$, $|\Psi_i\rangle, \,\, i=1,2$ reduce to $|+1\rangle\otimes |+1\rangle$ and $|+1\rangle\otimes |-1\rangle$  respectively. For $i=1$, it describes an asymmetric initial condition where one spin is aligned with their momentum and the other anti-aligned. In the second case, it corresponds to a symmetric initial state in which both spins are aligned with the momenta of the respective in-coming particle. At any other $\zeta$, the initial states are squeezed. As they are given here, these squeezed coherent states are not normalized. The proper normalization will be recovered in the computations of entropies shortly.

For starter, we can inspect analytically the Renyi-2 entropy after the scattering process. That is, we take the eikonal $\exp(i\chi)$ as the unitary evolution the initial states undergo and first construct the partial density matrix of the resulting final states by tracing out the spin degrees of freedom of particle 2. The Renyi-2 entropy is then computed straightforwardly from this partial density matrix. Similar to the scrambling case, we also see that the terms linear in $G_N$ cancel out here, regardless of initial states. The first non-trivial correction again comes in at $\mathcal O(G_N^2)$ and has the same power of $\hbar$ as the superclassical contributions in $\exp(i\chi)$. The Renyi-2 entropies after scattering with the two initial states can be calculated analytically. 

\begin{align}
&S_{|\Psi_1\rangle}^{(2)}= 4G^2_N(\text{sech}(2\zeta)+1)\left[|\mathbb A^{++}_{1,1}|^2+2|\mathbb A^{++}_{2,1}|^2+(\text{sech}(2\zeta)+1)(|\mathbb A^{++}_{1,2}|^2+2|\mathbb A^{++}_{2,2}|^2)\right], \,\,  \\
 &S_{|\Psi_2\rangle}^{(2)}= 4G^2_N(-\text{sech}(2\zeta)+1)\left[|\mathbb A^{++}_{1,1}|^2+2|\mathbb A^{++}_{2,1}|^2+(-\text{sech}(2\zeta)+1)(|\mathbb A^{++}_{1,2}|^2+2|\mathbb A^{++}_{2,2}|^2)\right] .
\end{align}

In the non-relativistic regime dictated by the conditions $m_{1,2}/p \gg 1/(|b|m_{1,2})$ and $\cosh\Theta \rightarrow 1$, we obtain the following leading-order behavior of the Renyi-2 entropies 

\begin{align}
\left. S_{|\Psi_1\rangle}^{(2)} \right|_{\text{non-rel}} =& {64\pi^4 G_N^2 m_1^2 m_2^2 \left(1+\text{sech}(2\zeta)\right) \over |b|^4 p^2 (m_1+m_2)^2} \,,\\
\left. S_{|\Psi_2\rangle}^{(2)} \right|_{\text{non-rel}} = &{64\pi^4 G_N^2 m_1^2 m_2^2 \left(1-\text{sech}(2\zeta)\right) \over |b|^4 p^2 (m_1+m_2)^2} \,,
\end{align}
where the subscript indicates which initial state the entropy is computed with. Here the leading behaviours are theory-agnostic for both initial states. The sub-leading terms however depend on the Wilson coefficients. We observe that the minimal coupling does not appear to be correlated with extremums in this case.

In the high energy regime defined by conditions $(p/m_{1,2})^2 \gg 1/(|b|m_{1,2})$ and $\cosh\Theta \gg 1$, we again divide our analysis into two scenarios according to the relative ratio of the competing scales $p/m_{1,2}$ and $|b|m_{1,2}$, as discussed in the previous section. 
\paragraph{Scenario I:} in addition to the two inequalities above, we also demand $|b| m_{1,2} \gg p/m_{1,2}$. To leading order, the Renyi-2 entropies read 
\begin{align}
\left. S_{|\Psi_1\rangle}^{(2)} \right|_{\text{high-energy}}=&{1024\pi^4 G_N^2 \left(1+\text{sech}(2\zeta)\right) \over |b|^4} \,,\\
\left. S_{|\Psi_2\rangle}^{(2)} \right|_{\text{high-energy}}=&{1024\pi^4 G_N^2 \left(1-\text{sech}(2\zeta)\right) \over |b|^4} \,.
\end{align}
We note that in this case the leading behaviour is theory independent and the first sub-leading terms, which we have not displayed, vanish at the minimal coupling.

\paragraph{Scenario II:} in the case $|b| m_{1,2}\ll p/m_{1,2}$ we have to leading order the following,
\begin{align}
\!\!\!\!\! \left. S_{|\Psi_1\rangle}^{(2)} \right|_{\text{high-energy}} =& {\frac{4608 G_N^2\pi^4p^4\left(1+\text{sech}(2\zeta)\right)^2}{b^8m_1^4m_2^4}(C_{1,2}-1)^2(C_{2,2}-1)^2} \,,\\
\!\!\!\!\! \left. S_{|\Psi_2\rangle}^{(2)} \right|_{\text{high-energy}} =&  {\frac{4608 G_N^2\pi^4p^4\left(1-\text{sech}(2\zeta)\right)^2}{b^8m_1^4m_2^4}(C_{1,2}-1)^2(C_{2,2}-1)^2}\,.
\end{align}
We observe that only in this case the leading behaviours have theory-dependences. The sub-leading also depends on the Wilson coefficients. Both the leading and the first sub-leading terms vanish at the minimal coupling.

\begin{figure}[tbp]
    \centering
    \includegraphics[width=.8\textwidth]{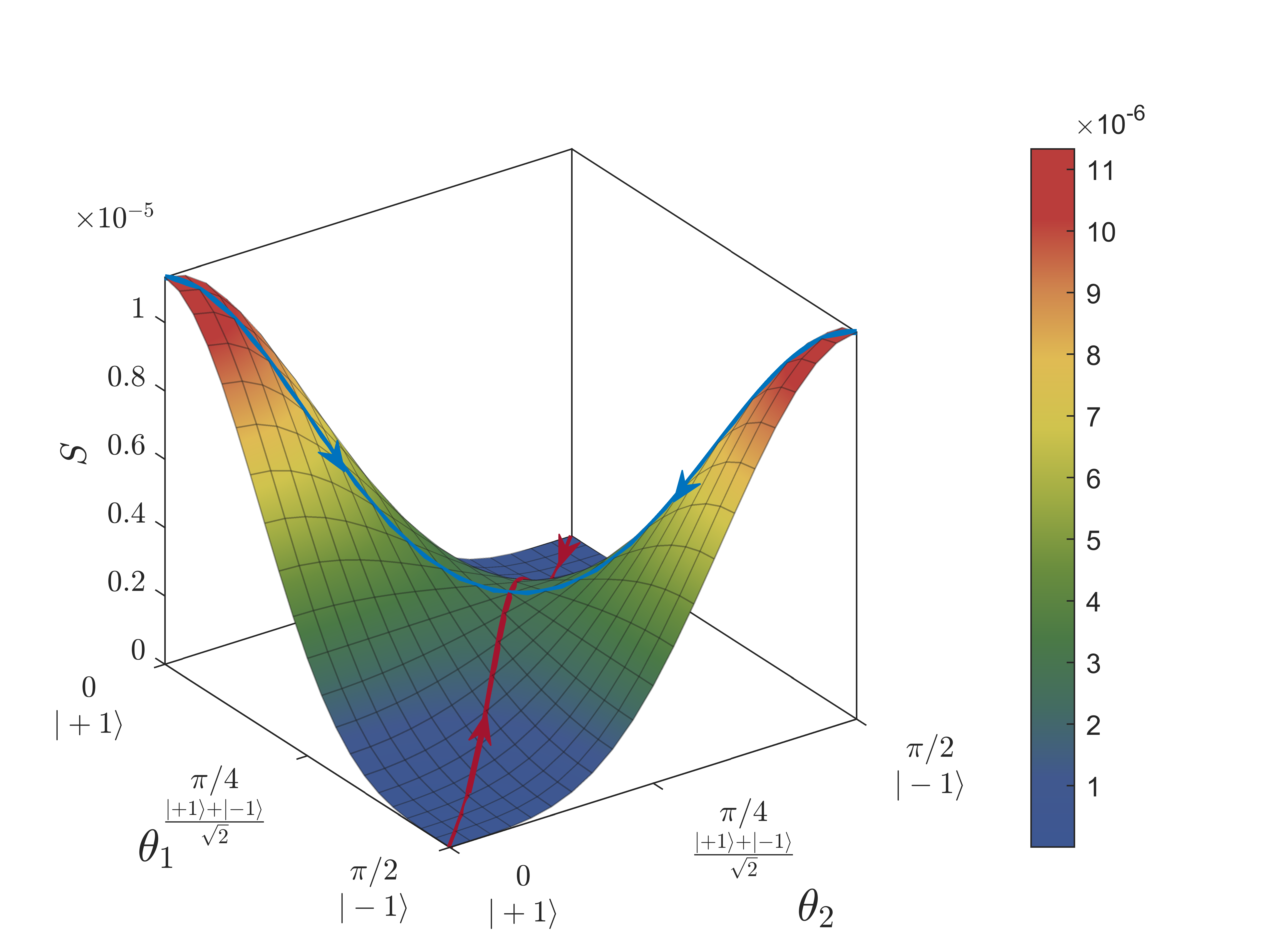}
    \caption{The entanglement entropy of the outcome states vs $\theta_1$ and $\theta_2$ with the initial states $(\text{cos}\theta_1|+1\rangle+\text{sin}\theta_1|-1\rangle)\otimes (\text{cos}\theta_2|+1\rangle+\text{sin}\theta_2|-1\rangle) $, $m_1=m_2=m$, $G_Nm^2=10^{-5}$, $|b|m=10$, $p/m=100$, $C_{1,2}=C_{2,2}=1.5$. This graph contains the configurations of squeezed states. The blue line shows when we squeeze {\it both} particles starting from parallel configurations ($|+1\rangle\otimes|+1\rangle$ or $|-1\rangle\otimes|-1\rangle$) to the middle point, the entanglement entropy decreases. The red line shows when we squeeze {\it both} particles starting from anti-parallel configurations ($|+1\rangle\otimes|-1\rangle$ or $|-1\rangle\otimes|+1\rangle$) to the middle point, the entanglement entropy increases.}
    \label{fig:Squeeze}
\end{figure}

One can see that the symmetric vs asymmetric configurations give {\it opposite } results. In the symmetric initial configuration, squeezing reduces final entanglement, making vanishing squeezing the minimum and most classical point, whereas in the anti-symmetric configuration, vanishing squeezing leads to {\it maximum} final entanglement after scattering, making the spin-coherent state the least classical configuration!
A global plot of the final entanglement entropy plotting against a pair of parameters that could smoothly interpolate between different squeezing and also relative orientations of the spins is shown in figure \ref{fig:Squeeze}. The colored lines correspond to symmetric and asymmetric relative orientations considered above.

While these are lower bounds of the entanglement entropy, exact numerical results can be obtained for example for spin-1, confirming the above statement. 
This is an interesting observation. While it is not entirely unexpected that final entanglement has dependence on details of the initial state, it is quite amusing that the spin coherent state can in fact be the most classical or the least classical initial state, depending on relative orientation between the two scattering particles. 
This opens the door to adjusting and fine-tuning initial states in future experiments to amplify the quantum effects of scattering event.

As noted before, spin coherent states are related to coherent states in harmonic oscillators in the large $j$ limit. It is thus of interest to explore the large $j$ limit, which we will defer for future work. We only comment here that the above computations do not appear to be very sensitive to the value of the spins and it is very conceivable that something similar would happen at larger values of $j$. 

\subsection{Mixed initial states}

By the same token, we are curious if any initial mixed state with the same expectation value of $J_x, J_y$ and $J_z$ can be entangled by different amount under gravitational scattering. 

As it is well known, entanglement entropy is a good measure of quantum entanglement if we work with pure quantum states. To deal with quantum entanglement of mixed states, it is more appropriate to work with the notion of negativity $\mathcal{N}$, which is defined as
\begin{align}
\mathcal{N}_A= ||\rho^{\Gamma_A}||_1, \qquad  || \mathcal{O}||_1 \equiv \textrm{tr} \sqrt{\mathcal{O}^\dag \mathcal{O}}
\end{align}
where $\rho$ is the density matrix of the full system, and $\Gamma_A$ denotes taking transpose of the density matrix only over degrees of freedom of the subsystem $A$. One then computes the norm of the resultant matrix.

We pick an initial mixed state that again preserves axial symmetry along the $z$ direction. For spin-$1$, consider a family of states that is parameterized as 
\begin{align}
|\psi\rangle_\phi=\sqrt{\frac{1}{2}}(|+1\rangle+e^{i\phi}|-1\rangle)\otimes |0\rangle, 
\end{align}
which has vanishing expectation value for spin angular momentum in all directions. 
We choose $\rho(\phi,\epsilon)=(1-\epsilon)|\psi\rangle_0\langle\psi|_0+\epsilon|\psi\rangle_\phi\langle\psi|_\phi$ to be the set of initial states. 
When $\phi=\pi$ and $\epsilon=1/2$, $\rho$ coincides with the equal mixture of the two coherent states $|+1\rangle$ and $|-1\rangle$.

\begin{figure}[tbp]
    \centering
    \includegraphics[width=.7\textwidth]{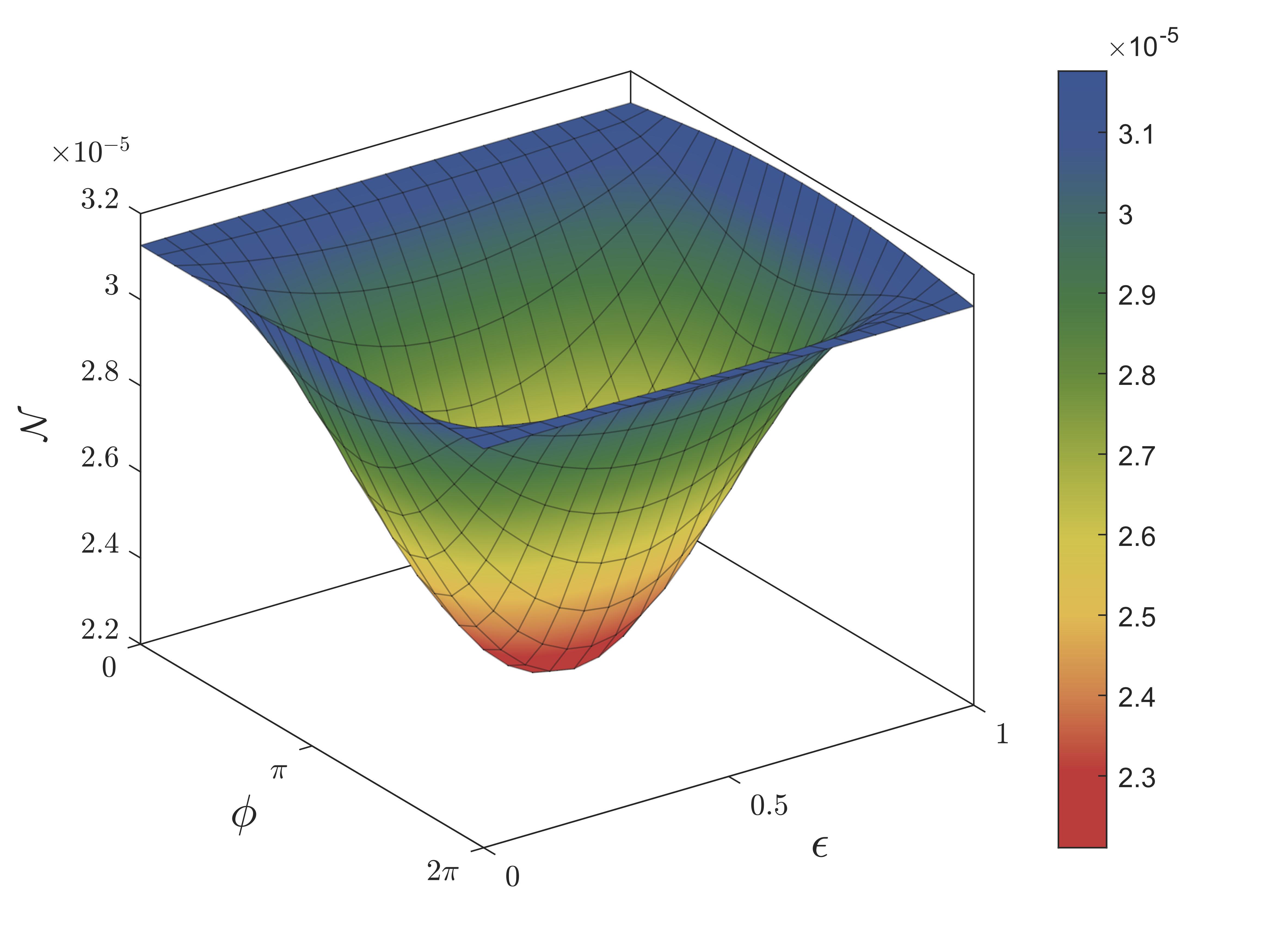}
    \caption{The negativity of the outcome states vs $\phi$ and $\epsilon$ with the initial states $\rho(\phi,\epsilon)$, $m_1=m_2=m$, $G_Nm^2=10^{-5}$, $|b|m=10$, $p/m=100$, $C_{1,2}=C_{2,2}=1$.}
    \label{fig:negativity}
\end{figure}

To quantify quantum entanglement in mixed states, we plot the negativity of the final states in Fig.~\ref{fig:negativity}. The quantum entanglement of the final states is minimal at $\phi=\pi$ and $\epsilon=1/2$.
 Mixed spin states that are closest to classical states are considered as those that can be expressed in terms of a positive combination of spin-coherent states, known as a P-rep  \cite{Sudarshan:1963ts, Glauber:1963tx, Giraud_2008}. i.e. 
\begin{align}
\rho_P = \int d\alpha  P(\alpha) |\alpha\rangle \langle \alpha |, \qquad P(\alpha) \ge 0 \,\,   \forall \alpha, \,\, \int d\alpha P(\alpha) = 1.
\end{align}
where $\alpha$ parametrizes the spin coherent states, and $d\alpha$ denotes the corresponding measure \cite{Sudarshan:1963ts, Glauber:1963tx, Giraud_2008}.
We note that the initial state that end up with minimal negativity is also the only P-rep state in the family of mixed states considered above.
Let us demonstrate that in detail in the following. 
Particle $2$ is in pure $|0\rangle$ state, so we consider only the density matrix of particle $1$. 
For spin-$1$, conditions for a density matrix to be a P-rep state has been studied in \cite{Giraud_2008}.  We borrow their results here. 
The density operator of a coherent state $|\mathbf{n}\rangle$ with unit vector $\mathbf{n}$ specifying a polarization vector in $\mathbb{R}^3$ can be written as \cite{Giraud_2008}
\begin{equation}
|\mathbf{n}\rangle\langle\mathbf{n}|=\frac{1}{3} \mathbf{1}_3+\frac{1}{2} \mathbf{n} \cdot \mathbf{J}+\frac{1}{2} \sum_{a, b=x, y, z}\left(n_a n_b-\frac{1}{3} \delta_{a b}\right) \frac{J_a J_b+J_b J_a}{2}
\end{equation} 
where $\mathbf{J}$ are the angular momentum matrices. A general density matrix is 
\begin{equation}
\rho=\frac{1}{3} \mathbf{1}_3+\frac{1}{2} \mathbf{u} \cdot \mathbf{J}+\frac{1}{2} \sum_{a, b=x, y, z}\left(W_{a b}-\frac{1}{3} \delta_{a b}\right) \frac{J_a J_b+J_b J_a}{2}.
\end{equation} 
For our choice, $\rho_1(\phi,\epsilon)=\begin{pmatrix} 1 & 0 & \delta \\ 0 & 0 & 0 \\ \delta^* & 0 & 1 \end{pmatrix}$ where $\delta=\frac{\epsilon}{2}e^{-i\phi}+\frac{1-\epsilon}{2}$, one can show that this gives $\mathbf{u}=(0,0,0)$ and $W=\frac{1}{2}\begin{pmatrix} \delta+\delta^* & i(\delta^*-\delta) & 0 \\ i(\delta^*-\delta) & -(\delta+\delta^*) & 0 \\ 0 & 0 & 2 \end{pmatrix}$.  

If a P-rep state  exists, then we can find $\lambda_i$ satsifying \cite{Giraud_2008}
\begin{equation} \label{eq:prepconditions}
\begin{gathered}
u_a = \sum_i \lambda_i n_a^{(i)} \\
W_{a b}=\sum_i \lambda_i n_a^{(i)} n_b^{(i)}
\end{gathered}
\end{equation}
where $\lambda_i$ is some real number satisfying $\lambda_i \ge 0$ and that $\sum_i \lambda_i=1$ \cite{Giraud_2008}. 
 Considering the diagonal components of $W$, we conclude that (\ref{eq:prepconditions}) can be satisfied for $W_{xx}=-W_{yy}=0$, and thus $n_x^{(i)}=n_y^{(i)}=0$.  
 If the $x,y$ components of $\mathbf{n}^{(i)}$ vanish,
 it immediately implies that $W_{xy} = W_{yx}=0$ from the conditions on $W$ in (\ref{eq:prepconditions}). 
 Therefore, $\rho_1(\phi,\epsilon)$ is P-rep only when $\delta=0$, which is the minimal point in the plot of negativity.

\section{Conclusions and Outlook}

We would like to revisit the question of the emergence of a sense of classicality in the context of particles interacting through gravity. 
Our study is two-folded. We first look into the scrambling power of a scattering event between two particles for given initial kinematics. 
We measure the power of scrambling power in spin space using the tripartite information computed from the eikonal amplitude, which is expected to diminish in a classical interaction -- since the tripartite information is measuring how the information gets delocalized through entanglement between widely separated particles generated under the scattering process. Our approach based on the tripartite information is independent of the initial spin polarization of the scattering particles, and depend only on the coupling of the particles to gravity, and the initial kinematics. Since we are interested in the classical limit {we are expanding the tripartite information to leading order in $G_N$ and the inverse impact parameter $1/|b|$. We  carefully isolate regions of kinematics so that the loop corrections remain sub-leading, ensuring the validity of our approximation.}
We found that the tripartite information, or more precisely an upper bound of the tripartite information based on the Renyi entropy, shows universal behaviour independent of theory both in the non-relativistic and a relativistic limit in which the impact parameter remains the most dominant scale. The scrambling power in these limits do diminish with increasing $|b|$. We note also that since they are theory agnostic, the minimal coupling does not play a visible role. 
Theory dependence comes in a somewhat different relativistic limit, in which the minimal coupling is singled out to lead to minimal scrambling, as is observed in \cite{Aoude:2020mlg,Chen:2021huj}, supporting at least in this limit that the minimal coupling is a special point that behaves most classically.
The exact tripartite information can be computed more readily for small spins, and we obtained explicit results for spin 1/2 and spin 1. Interestingly, they receive corrections logarithmic in $G_N$. There is a logorithmic boost to the scrambling power that would be otherwise missed in the Renyi entropy bound, and it bears some resemblance to the logarithmic corrections to the black hole entropy. We also computed the large spin limit of the bound of the tripartite information, which can be obtained analytically at minimal coupling. Perhaps somewhat as expected, it confirms the folklore that scrambling should diminish and thus the system appearing more classical as spins approach infinity while keeping $|s|/(b m) $ of order 1. 

On a different note, we explore the initial spin state dependence of entanglement generation.  We would like to explore the folklore that the coherent state is the most classically behaving state. In the case of harmonic oscillators, it was found that among squeezed coherent states that all saturate uncertainty inequalities, the amount of entanglement generation is minimal when there is no squeezing under generic local Hamiltonian evolution \cite{Zurek:1992mv}. In this paper we inspect squeezed spin coherent states as initial states and found that the state with vanishing squeezing happens to be an extremal point in the entanglement generation under gravitational scattering. However it could be either the minimal or the maximal depending on the initial relative orientation of the spins of the particles. While the result is not definitive selecting the usual spin coherent states, it suggests that careful choice of initial kinematics and orientations could amplify quantum signatures of a scattering event. 
We explored also a family of mixed initial states and used negativity as a measure of entanglement generation in the scattered final states. Interestingly, we found that in this case the minimally entangled final state do single out the so-called P-rep states which are considered as the best approximation of classical mixed spin states. 

To conclude, the study of scrambling and quantum entanglement generation provide some more intuition to the emergence of classicality in various corners of kinematic space and theory space, confirming a number of folklores.  In appropriate limits, we found universal behaviour independent of theories, which is a bonus to our study. It would be interesting to explore how other forms of interactions would show different universal behaviour. If so, it gives us an extra handle to detect and distinguish - perhaps even experimentally in the future - new fundamental forces. 
 
Our study scratches only the surface of the subject -- exploring the signature of quantum-ness/classicality of field theory in particular scattering amplitudes, through the lens of quantum information. For example, it would be interesting to compare our results with interactions via other spins, such as spin-$1$ and higher spin gauge fields and seek universal scaling behaviour. It would also be interesting to study the contribution to the amplitudes through the quantum remainder, which shows up starting from one loop -- to see if genuine quantum effects would produce qualitatively different quantum effects. Higher-point functions beyond four-point scattering would allow us to explore intricate entanglement pattern that gravity or other forces are capable of producing -- for example for 3 qubits there are exactly 2 different classes of entanglement not connected by LOCC -- the W class and the GHZ class \cite{Dur:2000zz}. 
There is a wealth of tools from the quantum information literature to be tapped into. We will leave the various questions we raised above to future work.

\appendix

\section*{Acknowledgements}
    We thank Aninda Sinha,  Si-Wei Zhong and Parthiv Halder for initial collaboration.  We also thank Yu-tin Huang for many helpful discussions. 
 LYH acknowledges the support of NSFC (Grant
No. 11922502, 11875111) and the Shanghai Municipal Science and Technology Major Project
(Shanghai Grant No.2019SHZDZX01)
    TW is supported by in part by the Special Research Assistantship of the Chinese Academy of Sciences and Humboldt Universit\"at zu Berlin. TW is also supported by the Fellowship of China Postdoctoral Science Foundation (No. 2022M713228), the Key Research Program of CAS, Grant No. XDPB15 and National Natural Science Foundation of China, Grants No. 11935013, No. 11947301, No. 12047502, and No. 12047503.

\appendix
\section{Renyi Entropy}\label{app.renyi}
We compute the Renyi entropy $S_{AC}$ explicitly. We note that the leading order of the reduced density matrix $\rho_{AC}$ is a projector and this leads to a discontinuity in $\text{tr}\rho^r_{AC}$ as $r\rightarrow 1$. Therefore for $r\geq 2$, we need to compute $\text{tr}\rho^r_{AC}$ and the normalisation $(\text{tr}\rho_{AC})^r$ separately.

Let us consider the normalisation first. Up to $\mathcal O(G_N^2)$, we have
\begin{align}
(\text{tr}\rho_{AC})^r  = & d_{S_2}^r \big(\text{tr} \mathbb P\big)^r + G_N \, r d_{S_2}^{r-1}\left(\text{tr} \Omega_1\right) \big(\text{tr} \mathbb P\big)^{r-1} \nonumber\\
& + G_N^2\, r d_{S_2}^{r-1} \left(\text{tr} \Omega_2^{(1)}+\text{tr} \Omega_2^{(2)} \right) \big(\text{tr} \mathbb P\big)^{r-1}  + G_N^2\, { r (r-1) \over 2} d_{S_2}^{r-2} \left(\text{tr} \Omega_1\right)^2 \big(\text{tr} \mathbb P\big)^{r-2} \nonumber\\ 
=& d_{S_1}^r d_{S_2}^r + G_N^2 \, r d_{S_1}^{r-1} d_{S_2}^{r-1} \left(\text{tr} \Omega_2^{(1)}+\text{tr} \Omega_2^{(2)} \right) \,,
\end{align}
where we denote $d_{s_a} = 2 s_a +1$ and $\left(\text{tr} \Omega_1\right) =0$ since $\text{tr}\Sigma^n_\pm =0$ and $\mathbb A_{00}^{nm}=0$ for any positive integers $n,m$. The two remaining trace terms $\text{tr}\Omega_2^{(1)}$ and $\text{tr}\Omega_2^{(2)}$ can be computed straightforwardly,
\begin{align}
\text{tr}\Omega_2^{(1)} &= \sum_{ \{n_a\} } \sum_{ \{I_a\} }{1\over 2}\left[ \mathbb A^{I_1 I_2}_{n_1 n_2} \mathbb A^{J_1 J_2}_{m_1 m_2} +\text{c.c.}\right] \text{tr}\big( O_{I_1 J_1}^{(n_1 m_1)}\big) \text{tr}\big( O_{I_2 J_2}^{(n_2 m_2)} \big)  \,,\\
\text{tr}\Omega_2^{(1)} &= \sum_{ \{n_a\} } \sum_{ \{I_a\} } \mathbb A^{I_1 I_2}_{n_1 n_2} \left(\mathbb A^{J_1 J_2}_{m_1 m_2}\right)^* \text{tr}\big( O_{I_1 J_1}^{(n_1 m_1)}\big) \text{tr}\big( O_{I_2 J_2}^{(n_2 m_2)} \big)  \,.
\end{align}
Now we compute $\text{tr}\rho^r_{AC}$ up to $\mathcal O(G_N^2)$. For $r\geqslant2$, we have
\begin{align}\label{eq:trRhoACr}
\text{tr}\rho^r_{AC} = &\, \text{tr}\left(\mathbb P^r\right) + G_N\,r\, \text{tr}\left( \Omega_1 \mathbb P^{r-1} \right) +G_N^2\, r\, \text{tr}\left[ \big(\Omega_2^{(1)} + \Omega_2^{(2)}\big) \mathbb P^{r-1}\right] \nonumber\\
& + G_N^2 {r\over 2}\sum_{1 < p  \leqslant r} \text{tr}\left[ \Omega_1 \mathbb P^{p-2} \Omega_1 \mathbb P^{r-p}\right]\,.
\end{align}
Recall that $\mathbb P^{r} = d_{s_1}^{r-1} d_{s_2}^{r-1} \mathbb P$ and therefore the first line above can be calculated immediately,
\begin{align}
\text{tr}\left(\mathbb P^r\right) = d_{S_1}^r d_{S_2}^r\,,~~~~
\text{tr}\left( \Omega \mathbb P^{r-1} \right) = d_{S_1}^{r-2} d_{S_2}^{r-2} \text{tr} \left( \Omega \mathbb P\right)\,,
\end{align}
where $\Omega = \{ \Omega_1, \Omega_2^{(1)},\Omega_2^{(2)} \}$\,. These traces read
\begin{align}
\text{tr} \big( \Omega_1 \mathbb P\big) &= d_{s_1} d_{s_2} \sum_{ \{ n_a\} } \sum_{ \{ I_a \} } \left[ \mathbb A_{n_1 n_2}^{I_1 I_2} + \text{c.c.} \right] \text{tr}\big( O_{I_1}^{(n_1)}\big) \text{tr}\big( O_{I_2}^{(n_2)}\big) \,,\\
\text{tr} \big( \Omega_2^{(1)} \mathbb P\big) &= {d_{s_1} d_{s_2} \over 2} \sum_{ \{n_a \} } \sum_{ \{I_a\} } \left[ \mathbb A_{n_1 n_2 }^{I_1 I_2} \mathbb A_{m_1 m_2}^{J_1 J_2}+\text{c.c.}\right] \text{tr} \big( O_{I_1 J_1}^{(n_1 m_1)}\big) \text{tr}\big( O_{I_2 J_2}^{(n_2 m_2)}\big)\,,\\
\text{tr} \big( \Omega_2^{(1)} \mathbb P\big)  &= d_{s_2} \sum_{\{n_a \} } \sum_{\{ I_a \} } \mathbb A_{n_1 n_2 }^{I_1 I_2} \left( \mathbb A_{m_1 m_2}^{J_1 J_2} \right)^* \text{tr} \big( O_{I_1}^{(n_1)}\big) \text{tr} \big( O_{J_1}^{(m_1)}\big) \text{tr}\big( O_{I_2 J_2}^{(n_2 m_2)}\big)\,.
\end{align}
Similar to the normalisation, we have $\text{tr} \big( \Omega_1 \mathbb P\big)=0$. The second line of (\ref{eq:trRhoACr}) involves several structures. For $r=2$, there is only one term in the sum, namely $\sum_{1 < p  \leqslant r} \text{tr}\left[ \Omega_1 \mathbb P^{p-2} \Omega_1 \mathbb P^{r-p}\right] = \text{tr}\Omega_1^2$ and we have 
\begin{align} 
\text{tr}\Omega_1^2 = 2 d_{s_1}\sum_{\{n_a, I_a\} }   \mathbb A_{n_1 n_2}^{I_1 I_2} \left(\mathbb A_{m_1 m_2}^{J_1 J_2}\right)^* \text{tr}\big( O_{I_1 J_1}^{(n_1 m_1)}\big)  \text{tr}\big( O_{I_2}^{(n_2)}\big) \text{tr}\big( O_{J_2}^{(m_2)}\big) \,.
\end{align}
For $r=3$, there is only one structure $\sum_{1 < p  \leqslant r} \text{tr}\left[ \Omega_1 \mathbb P^{p-2} \Omega_1 \mathbb P^{r-p}\right] = 2\text{tr}\big( \Omega_1^2 \mathbb P\big)$ and 
\begin{align}
\text{tr}\big( \Omega_1^2 \mathbb P\big)
= d_{s_1}^2 d_{s_2} \sum_{\{n_a,I_a\} }  \mathbb A_{n_1 n_2}^{I_1 I_2} \left(\mathbb A_{m_1 m_2}^{J_1 J_2}\right)^* \text{tr}\big( O_{I_1 J_1}^{(n_1 m_1)}\big)  \text{tr}\big( O_{I_2}^{(n_2)}\big) \text{tr}\big( O_{J_2}^{(m_2)}\big)
\end{align}
For $r\geqslant4$, there are two structures as follows,
\begin{align}
\sum_{1 < p  \leqslant r} \text{tr}\left[ \Omega_1 \mathbb P^{p-2} \Omega_1 \mathbb P^{r-p}\right] = 2 d^{r-3}_{s_1} d_{s_2}^{r-3} \text{tr} \big( \Omega_1^2 \mathbb P\big) + (r-3) d^{r-4}_{S_1} d_{S_2}^{r-4} \text{tr} \big( \Omega_1 \mathbb P \Omega_1 \mathbb P\big)\,.
\end{align}
It is straightforward to see that $\text{tr} \big( \Omega_1 \mathbb P \Omega_1 \mathbb P\big)$ vanishes in the same way as the terms linear in $G_N$. 
Hence the second line of (\ref{eq:trRhoACr}) takes the following expression for $r\geqslant 2$,
\begin{align}
& {r\over 2} \sum_{1 < p  \leqslant r} \text{tr}\left[ \Omega_1 \mathbb P^{p-2} \Omega_1 \mathbb P^{r-p}\right] \nonumber\\
=\,& r d_{s_1}^{r-1} d_{s_2}^{r-2}  \sum_{\{n_a, I_a\} }   \mathbb A_{n_1 n_2}^{I_1 I_2} \left(\mathbb A_{m_1 m_2}^{J_1 J_2}\right)^* \text{tr}\big( O_{I_1 J_1}^{(n_1 m_1)}\big)  \text{tr}\big( O_{I_2}^{(n_2)}\big) \text{tr}\big( O_{J_2}^{(m_2)}\big)
\end{align}

The Renyi entropy can now be readily computed as follows,
\begin{align}
{\text{tr}\left( \rho^r_{AC}\right) \over \left(\text{tr} \rho_{AC}\right)^r} &= \frac{ d_{s_1}^r d_{s_2}^r + G_N^2 r d_{S_1}^{r-2} d_{S_2}^{r-2} \left[ \text{tr}\big(\Omega_2^{(1)} \mathbb P\big) +\text{tr}\big(\Omega_2^{(2)} \mathbb P\big)\right] +  { G_N^2 r\over 2} \sum\limits_p\text{tr}\left[ \Omega_1 \mathbb P^{p-2} \Omega_1 \mathbb P^{r-p}\right]}{ d_{S_1}^r d_{S_2}^r  + G_N^2 r d_{S_1}^{r-1} d_{S_2}^{r-1} \left[ \text{tr}\Omega_2^{(1)} +\text{tr}\Omega_2^{(2)}\right]} \nonumber\\
&= 1+{G_N^2 r \over d_{s_1}^2 d_{s_2}^2}\left[ \text{tr}\big( \Omega_2^{(2)} \mathbb P\big) - d_{S_1} d_{S_2} \text{tr}\Omega_2^{(2)} \right] + { G_N^2 r\over 2 d_{s_1}^r d_{s_2}^r} \sum\limits_p\text{tr}\left[ \Omega_1 \mathbb P^{p-2} \Omega_1 \mathbb P^{r-p}\right] \nonumber\\
&= 1- {G_N^2 r \over d_{s_1} d_{s_2}}\sum_{n_a=1}^{2S_a} \sum_{I_a = \pm} \mathbb A_{n_1 n_2}^{I_1 I_2} \big(\mathbb A_{n_1 n_2}^{\bar I_1 \bar I_2} \big)^* t_1^{(n_1)} t_2^{(n_2)}\,,
\end{align}
where in the second line we have used $\text{tr}\big(\Omega_2^{(1)}\mathbb P\big) = d_{S_1} d_{S_2} \text{tr}\Omega_2^{(1)}$ and in the third line we have plugged in the explicit expressions for the remaining traces. Taking into account that $\text{tr}\big(\Sigma_\pm^n\big) =0$ and $\mathbb A_{nm}^{00}=0$ for any positive integers $n,m$, it is straightforward to find the final expression in which $\bar I_a = \mp$ for $I_a=\pm$.
    
\bibliography{ref}
\bibliographystyle{JHEP} 
\end{document}